\documentclass{article}
\usepackage{amsmath,amssymb,amsfonts}
\usepackage{graphics,subfigure}
\usepackage{epsfig}
\usepackage[dvipdfm]{hyperref}

\renewcommand{\eqref}[1]{Eq.~(\ref{#1})}

\newcommand{\avq}{\langle q \rangle}

\newcommand{\averageq}[1]{\langle q^{#1} \rangle}

\newenvironment{eqns}[1]{\begin{subequations}\label{#1}\begin{eqnarray}}{\end{eqnarray}\end{subequations}\ignorespacesafterend}

\title{Ising model for distribution networks\footnote{Dedicated to Professor Dr. David Sherrington on the occasion of his 70th birthday.} }

 \author{\small H. Hooyberghs$^{\rm a}$  $^{\ast}$, S. Van Lombeek$^{\rm a}$, C. Giuraniuc$^{\rm a,b}$,
         B. Van Schaeybroeck$^{\rm a,c}$ and  J. O. Indekeu$^{\rm a}$\\\vspace{6pt} \small
\hspace{-1cm}$^{\rm a}$ Instituut voor Theoretische Fysica, Katholieke Universiteit Leuven, Celestijnenlaan~200D, 3001 Heverlee, Belgium\\
\small $^{\rm b}$ Institute of Medical Sciences, University of Aberdeen, Foresterhill, Aberdeen~AB25~ 2ZD, UK\\
\small $^{\rm c}$ Koninklijk Meteorologisch Instituut (KMI), Ringlaan 3, 1180 Brussels, Belgium
}

\date{\small \today}

\begin{document}
 \maketitle

\begin{abstract} An elementary Ising spin model is proposed for demonstrating cascading failures (breakdowns, blackouts, collapses, avalanches, ...) that can occur in realistic networks for distribution and delivery by suppliers to consumers. A ferromagnetic Hamiltonian with quenched random fields results from policies that maximize the gap between demand and delivery. Such policies can arise in a competitive market where firms artificially create new demand, or in a solidary environment where too high a demand cannot reasonably be met. Network failure in the context of a policy of solidarity is possible when an initially active state becomes metastable and decays to a stable inactive state. We explore the characteristics of the demand and delivery, as well as the topological properties, which make the distribution network susceptible of failure. An effective temperature is defined, which governs the strength of the activity fluctuations which can induce a collapse. Numerical results, obtained by Monte Carlo simulations of the model on (mainly) scale-free networks, are supplemented with analytic mean-field approximations to the geometrical random field fluctuations and the thermal spin fluctuations. The role of hubs versus poorly connected nodes in initiating the breakdown of network activity is illustrated and related to model parameters.
\newline
\newline
Version: \today\\
\end{abstract}
{{\sc keywords: Ising model, random fields, complex scale-free distribution network, failure, blackout, breakdown, collapse, avalanche, cascade}}

\section{Motivation}
Statistical physics studies of cooperative phenomena on random graphs have attracted a lot of new attention and undergone impressive new development since it has become clear that many real-life interconnected structures are well approximated by random scale-free networks \cite{barab1,strogatz,newmanrev,dorogov,heiko}. One can say that a paradigm shift is occurring from studies of models for critical phenomena on (Bravais) lattices to studies in which such models are defined on random networks. To some extent this paradigm shift resembles that from Euclidean geometry to fractal geometry, in the modeling of various natural phenomena, but scale-free networks are specific in that they are characterized by the presence of a small but important set of {\em hubs}. The hubs are highly connected nodes which typically have a large influence on the operation and coherence of the structure.

When we are concerned with the distribution of electricity, or the production and sale of material goods by commercial centers to consumers, or the delivery of the daily mail ..., we can envisage distribution centers as nodes and consumers as links on a network. The consumers have certain demands and the distribution centers certain deliverables. The distribution centers can be active or inactive depending on internal conditions and on external criteria such as the demands of consumers that are linked to them and the status of other nearby centers. The occurrence of hubs in such networks is rather common (main providers, supermarkets, ...).

We propose a model that can capture three types of common distribution policies in realistic trade environments by implementing an appropriate Ising spin Hamiltonian on a random network, mainly of scale-free type, but we also consider networks with a scale. We present the three policies in a logical order, from obviously cooperative to rather ambivalent, and our main goal in this paper will be to demonstrate how a policy in the latter category can cause a breakdown of the distribution network. The first policy is one that is {\em guided by demand}. The distribution center aims at satisfying the consumer demand as closely as possible and will strive to adjust its activity to that of neighbouring centers so that the difference between demand and delivery is minimal on the links that lead to those centers. Note that every consumer (link) can be served by two centers (the adjacent nodes) in this model. This is not unusual. Many consumers rely on an alternative option in case their normal provider is not available. A distribution center $i$ is active or {\em up} when $\sigma_i=1$ and inactive or {\em down} when $\sigma_i=-1$. The {\em demand-guided policy} is characteristic of a market in which {\em compensation or complementarity is more important than competition}. Many physical, chemical and biological systems are equipped with similar {\em negative feedback} mechanisms, rendering operation stable under small enough perturbations.

The second distribution policy is concerned with a {\em competitive market} guided by product quality and quantity. In this policy production or distribution centers strive to maximize their activity in order to get {\em their} deliverables consumed rather than those of their rivals. The centers aim at manipulating consumers by {\em creating new demand}, preferably in situations where the existing demand is low. In this {\em product-guided policy} the centers tend to become active especially when their neighbours are already satisfying the consumer demand. This policy can therefore be mimicked by maximizing the difference between supply and demand, which is opposite to what we postulated in the demand-guided policy described before.

Thirdly, we wish to capture a {\em policy of solidarity} rather than competition, in which providers display a voluntary or forced shutdown in situations where {\em too high a demand cannot reasonably be met}. As long as all centers are active there are no problems, but as soon as some become inactive, the burden of satisfying the high consumer demand can become too heavy to carry. The centers can be unwilling or plainly incapable of rectifying the drop in supply, and increasingly so as time progresses and more neighbouring centers fail. This is a pronounced {\em positive feedback} mechanism which can lead to blackouts in power stations or certain kinds of strikes (when workers are unable or unwilling to do more than their peers). A similar conduct may be observed, for example, among partners in projects. Here we assume that a link is a project shared between adjacent nodes, which can model persons or companies.  Companies tend to close or, equivalently, persons tend to stop doing their job, when their neighbours no longer invest in a joint project (i.e., when neighbouring sites are down). Like in the (second) competitive policy we discussed, in this (third) solidarity policy the difference between supply and demand is again maximized, but this time by the supply dropping to zero.

In order to integrate these three policies in an Ising model Hamiltonian description we define, besides the spin states associated with the distribution centers, the supply and demand variables on the links. For the purpose of this Motivation, we focus on the simplest version of the model, but still sufficiently equipped to bring out the essential physics. In later Sections a more refined approach will be outlined. For the time being we assume that each distribution center, when active, delivers a fixed total amount (set to unity) to all of the links attached, divided equally over all links. For the link between nodes $i$ and $j$ with respective degrees $q_i$ and $q_j$, the delivery thus equals
\begin{equation}
\mathcal{L}_{ij}({\sigma_i,\sigma_j})
=\frac{1}{q_i}\left(\frac{\sigma_i+1}{2}\right)+\frac{1}{q_j}
\left(\frac{\sigma_j+1}{2}\right)
\end{equation}
Treating, provisionally, the demand as a uniform constant $\mathcal{D}$ throughout the network, we arrive according to the policies described above at an energy per link which can be written as
\begin{equation}
E_{ij}({\sigma_i,\sigma_j})
=-J [ \mathcal{L}_{ij}({\sigma_i,\sigma_j})- \mathcal{D}  ]^2,
\end{equation}
where the nearest-neighbour coupling $J$ is {\em negative} for a {\em demand-guided} policy and {\em positive} for the policies driven by {\em competition or solidarity}. The Hamiltonian of the Ising distribution network is then the sum of these energies over all links. We obtain the {\em spin Hamiltonian} of a given realization of a network with $N$ nodes,
\begin{equation}
\mathcal{H}(\mathbf{\{\sigma\}})=\sum_{<
ij>}E_{ij}({\sigma_i, \sigma_j})=-\frac{J}{2}\sum_{<
ij>}  \frac{\sigma_i \sigma_j}{
q_i q_j} - \sum_{i=1}^{N}H_i \sigma_{i} + const.
\label{Hami}
\end{equation}
where $<ij>$ denotes nearest-neighbour pairs and the {\em quenched random field} $H_i$ acting on the spin at node $i$ is given by
\begin{equation}
H_{i}=\frac{J}{2q_i} \left ( 1 + \sum_{j=1}^{q_i}\frac{1}{q_j}\right ) - J\mathcal{D}
\end{equation}
The spin-independent constant (last term in \eqref{Hami}) is irrelevant for our purposes and will henceforth be omitted.

The interpretation of the quenched random field is fairly transparent.
For $J<0$ (demand-guided policy) a distribution center experiences a bias to become active when the total demand of the attached links, $q_i \mathcal{D}$,  exceeds the total {\em average delivery} to these links, $(1+\sum_j q_j^{-1})/2$, where the average is taken over the spin values, i.e., over active and inactive states. For $J>0$ the spin bias is opposite. A distribution center tends to become inactive when the demand is high.

The nearest-neighbour interaction, $J/(2q_iq_j)$, is also a quenched random variable, the sign of which is equal to that of $J$. Note that this degree-dependent  interaction is reminiscent of that in the Special Attention Network model \cite{indekeu1}, in the sense that {\em high degree is compensated by weak interaction}. In the distribution model this means that the mutual influence between neighbouring centers is proportional to the product of their deliveries to the link that connects them, or to some power of this product. We conclude that we are dealing with a random-field (anti-)ferromagnetic Ising model on a network for $(J<0)$ $J>0$. The ground state, which minimizes the total energy, is in general non-trivial. In addition, ``thermal" noise is present, due to occasional maintenance shutdowns of centers or fluctuations of their activity caused by other internal factors. We therefore consider the model at a finite temperature $T$ to allow for these more or less random perturbations. The ratio $k_BT/J$ is a measure of their importance. Preliminary results of Monte Carlo simulations of the model at hand were reported by Giuraniuc \cite{thesisGiuraniuc}.

\section{General formulation of the distribution model}
\subsection{Construction of a general Hamiltonian}
Consider a static scale-free network with $N$ nodes. The normalized distribution of the degrees, the number of links attached to each node, $P(q)$, with $\sum_q P(q) = 1$, is assumed to follow a decreasing power-law  $P(q) \propto q^{-\gamma}$ for large $q$. The topological exponent $\gamma$ usually lies in the regime $2<\gamma<3$ for real-life networks \cite{newmanrev}. Each node $i$ represents a distribution center
which can be either active ($\sigma_i=+1$) or inactive
($\sigma_i=-1$). A link between nodes $i$ and $j$, denoted by $<
ij>$, is a \textit{consumer} or a region across which the products are
distributed. Contrary to the simplified model proposed in the Motivation, the nodes can produce unequal amounts of goods. We assume that the total delivery depends on the degree of the supplier. Therefore, the delivery  $\mathcal{L}_{ij}({{\sigma_i,\sigma_j}})$ to a link $<ij>$ is defined as
\begin{equation}
\mathcal{L}_{ij}({\text{{$\sigma_i,\sigma_j$}}})
=\frac{1}{q_i^\mu}\left(\frac{\sigma_i+1}{2}\right)+\frac{1}{q_j^\mu}
\left(\frac{\sigma_j+1}{2}\right),
\end{equation}
where $q_i$ and $q_j$ are the degrees of the nodes linked by $<ij>$. The exponent $\mu$ controls the total production of a supplier. When node $i$ is active or $\sigma_i=+1$, supplier $i$ furnishes $1/q_i^{\mu}$ products to each of its $q_i$ consumers. The total delivery of the active node is thus $q_i^{1-\mu}$. The case $\mu = 1$ corresponds to the special case in the Motivation in which all active nodes provide the same total supply. Consumers attached to highly-connected  nodes then receive a smaller amount. In general, the hubs deliver less goods to a single consumer for $\mu >0$.
For $\mu = 0$, every consumer would be receiving the same amount of products in an active state, which also implies that hubs have to deliver more goods in total.  Finally, $\mu <0$ corresponds to the situation in which the highly-connected suppliers can deliver more goods to each consumer.

Following the reasoning introduced in the Motivation, every consumer has a demand $\mathcal{D}$. We assume that the demand of link  $<ij>$ may depend on $q_i$ and $q_j$. The energy per link is now given by
\begin{equation}
E_{ij}(\sigma_i,\sigma_j)=-J\left [\mathcal{L}_{ij}({\sigma_i,\sigma_j})-\mathcal{D}_{ij}
\right ]^2.
\end{equation}
The total Hamiltonian becomes
\begin{eqnarray} \label{hamiltoniaan}
\mathcal{H}({\{\sigma\}})&=&\sum_{<
ij>}E_{ij}(\sigma_i,\sigma_j)=-\sum_{i=1}^{N}
 (H_i+I_i(\{\sigma\}))\sigma_{i}\\
 \text{with}&\quad\quad& H_{i}=
\sum_{j=1}^{q_i}\frac{J}{2q_i^{\mu}}\left(\frac{1}{q_i^\mu}+\frac{1}
{q_j^\mu}-2 \mathcal{D}_{ij}\right)\quad\quad
 \text{and}\quad\quad I_{i}(\{\sigma\})= \frac{J}{2q_i^\mu}
\sum_{j=1}^{q_i}\frac{\sigma_j}{q_j^\mu},\nonumber
\end{eqnarray}
up to constant (spin-independent) terms, which are neglected.
The sums over $j$ run over the $q_i$ neighbours of node $i$. This  short reasoning leads to two fields applied to
each node $i$. The \textit{quenched random field} $H_i$ is
inherent in the network and the \textit{interaction
field} $I_i$ originates from the interactions with spins linked to $\sigma_i$.

In this Paper we focus mainly on the third policy introduced in the Motivation. We describe avalanches in distribution networks which occur due to the {\em solidarity among suppliers}. The interaction constant $J$ is thus positive. Note that avalanches in anti-ferromagnetic spin systems with a uniform external field on complex networks have been studied in Ref.~\cite{malarz}. Avalanches in distribution models were also studied using sand piles on networks \cite{goh} and in various models based on critical loads \cite{sachtjen,watts,motter,crucitti}. Recently, also models studying percolation in interdependent networks  were used \cite{buldyrev}.

\subsection{The demand function}
Finally, we propose a suitable
demand function. The demand of a link $<ij>$ between nodes $i$ and $j$ is a combination of a link-dependent
and a homogeneous part:
\begin{align}\label{demand}
\mathcal{D}_{ij}=\frac{a}{2}\left(\frac{1}{q_i^\mu}
+\frac{1}{q_j^\mu}\right) +\frac{b}{2}\frac{\langle
q^{1-\mu}\rangle}{\langle q\rangle},
 \end{align}
where $a(>0)$ and $b$ are real constants. The notation $\langle\cdot \rangle$ denotes an average  over the degree distribution $P(q)$. The
first term, the \textit{supply-adjusted demand} is a consequence of the
form of the fields $H_i$ in \eqref{hamiltoniaan}. The parameter $a$ controls which fraction of the normal
capacity is demanded by the consumers on the link. Each link $<ij>$ also features an intrinsic uniform demand, independent of the deliveries of the nodes $i$ and $j$. The second term, regulated by the constant $b$, provides such a \textit{global demand}. The onset of a homogeneous consumer demand
($b\neq 0$) results in a total field on the network which has
the same order of magnitude  as the interaction field in an active state averaged over an ensemble of different networks.
Note that in certain systems individual consumers can deliver goods {\em to} the market. For instance, individual households can contribute to the power grid with the output of their own solar cells. Negative values of $b$ are therefore not excluded.

The total field acting on node $i$ is now given by
\begin{align}\label{totaalveld}
H_{i}+I_{i}(\{\sigma\})=\sum_{j=1}^{q_i}\frac{J}{2q_i^
{\mu}}\left[(1-a)\left(\frac{1}{q_i^\mu}+\frac{1}{q_j^\mu}\right)
-b\frac{\langle q^{1-\mu}\rangle }{\langle q\rangle}
+\frac{\sigma_j}{q_j^\mu}\right].
\end{align}
Note that only the last term depends on the interaction with the spins on the neighbouring sites. It will play a prominent role in the cascading effect.  For general values of $\mu$, both the interaction and the quenched random field applied on node $i$ will depend on the connectivity of node $i$ and on the connectivity of its neighbours. However, in some cases simplifications can be made. If $\mu = 0$, the parameters $a$ and $b$ can be replaced by a single parameter $a_0 = a+ b/2$. The total field is then simplified to
\begin{equation}
 H_i + I_i(\{\sigma\}) = Jq_i \left(1-a_0\right) + \frac{J}{2}\sum_{j=1}^{q_i}\sigma_j;\hspace{0.5cm} \textrm{for} \quad\mu = 0. \label{richtingkust}
\end{equation}
Both the interaction field and the quenched random field acting on a node are then independent of the degrees of the neighbouring sites. For arbitrary $\mu$ the quenched random field is independent of the connectivity of the neighbouring nodes for $a= 1$. Note finally that for $a=1$ and $b=0$ the demand just matches the average supply (the average being taken over active and inactive providers) and the quenched random field vanishes, i.e., $H_i =0$, $\forall i$. An Ising model with connectivity-dependent coupling constants in the absence of an external field was studied by Giuraniuc {\em et al.}~\cite{giuraniuc1}. It was shown to be equivalent to a model with homogeneous couplings and with a modified topological exponent $\gamma' = (\gamma - \mu)/(1-\mu)$.

The quenched random and interaction fields featured in \eqref{totaalveld} depend on the local structure of each network.  Handy analytic approximations can be obtained by performing an ensemble average over different network realizations with the same degree distribution.
In a first step, we  average over the degree of the neighbours to obtain the average field applied on node $i$. Let $P^{(i)}_n(q_j)$ denote the probability that the neighbour of node $i$ has connectivity $q_j$.  We assume that the network is uncorrelated. The distribution  $P^{(i)}_n(q_j)$ is then independent of node $i$ and related to the degree distribution $P(q_j)$ by the relation $P_n(q_j) = q_jP(q_j)/\avq$.
Therefore we define the average over the degrees of the
neighbouring sites of node $i$ as follows:
\begin{align}\label{averagen}
\ll\cdot\gg_i\:=\prod_{j=1}^{q_i}\sum_{q_j}\frac{q_jP(q_j)}{\langle
q\rangle}.
\end{align}
This type of averaging is a ``topological" mean-field approximation.
Simultaneously, we may in some situations wish to apply a ``thermal" mean-field approach to the thermally fluctuating spin variables and replace the actual spins of the neighbours by a mean spin. If we furthermore assume that the mean spin is homogeneous throughout the network, the average quenched random and interaction fields acting on node $i$ in the mean-field approach are
\begin{eqns}{velden}
\ll
H_i\gg_i&=&Jq_i^{1-2\mu}\frac{1-a}{2}+Jq_i^{1-\mu}\frac{\langle
q^{1-\mu}\rangle}
{2\langle q\rangle}\left(1-a-b\right),\\
\ll I_i(\{\sigma\})\gg_i&=&Jq_i^{1-\mu}\frac{\langle
q^{1-\mu}\rangle}{2\langle q\rangle} \sigma_{av},
\end{eqns}
with $\sigma_{av}$ the average spin of a node. Deviations
from these averages are important for a network in which the
degree distribution $P(q)$ possesses diverging moments. For instance, in a
scale-free network with topological exponent $\gamma\leq 3$, the second moment diverges. Note that the average quenched random and interaction fields  fields on a node diverge for $\mu \leq 2-\gamma$. Therefore we will henceforth limit our attention to the case $\mu > 2 - \gamma$. Since we only consider scale-free networks with a finite mean degree, i.e., with $\gamma > 2$, this limitation is only important for $ \mu < 0$.

In a second step, we average over the degrees of the nodes using the degree distribution $P(q)$. The average Hamiltonian $\langle\mathcal{H}(\sigma_{av})\rangle$ is then given by
\begin{equation}
 \langle\mathcal{H}(\sigma_{av})\rangle = \sum_i \sum_{q_i} P(q_i) \sigma_{av}\ll
H_i+ I_i(\sigma_{av})\gg_i.
\end{equation}
Inserting both members of \eqref{velden} we obtain
\begin{equation}\label{meanham}
  \langle\mathcal{H}(\sigma_{av})\rangle=-N J\sigma_{av} \left(\averageq{1-2\mu}\frac{1-a}{2} + \frac{\averageq{1-\mu}^2}{2\avq}\left(1-a-b + \sigma_{av}\right) \right).
\end{equation}
\eqref{meanham} provides a mean-field expression for the Hamiltonian averaged over different realizations of the same network.

\section{Requirements and favourable circumstances for a network collapse}
For the study of a collapsing network, driven by competition or solidarity ($J>0$), we exploit the {\em ferromagnetic}
phase which appears in the Ising model at sufficiently low
temperatures and for sufficiently weak random-field fluctuations. Recall that the ground state ($T=0$) in the absence of fields consists of two degenerate configurations: the \textit{active} network in which $\forall$ $i$,
$\sigma_i=+1$, and the \textit{inactive} network in which $\forall$ $i$, $\sigma_i=-1$. By applying a uniform bulk field, the degeneracy is
lifted and a unique equilibrium state emerges. Nevertheless, the network can reside for a long time in the oppositely magnetized ``metastable"
state if the fields and (thermal) spin fluctuations are too small to
trigger the collapse to the equilibrium state.

Metastability is defined {\em dynamically} in our context as partial stability with respect to certain perturbations, in contrast with absolute stability (with respect to any perturbation). In our model, in the presence of random fields of microscopic origin, which are inhomogeneous and
regulated by the parameters $a$, $b$ and $\mu$, the ground state may be non-trivial and will coincide with the all-up or all-down states only under certain conditions on the random fields. Therefore, considering metastable states, we need to distinguish between states that decay to a ferromagnetic state or evolve to a ``glassy" state. We will be interested mostly in the ferromagnetic state because in a glassy state the network is still more or less active.

Thus, in our study of distribution networks driven by solidarity, we will mainly focus on the decay of the all-up state (active network) to an essentially all-down state (inactive network). When the global consumer demand (modeled by $b$) is increased, the metastable states we study by means of a suitable spin-flip dynamics, mimic the behavior of certain realistic
systems. For example, from 1988 to 1998, US electricity demands increased by nearly 30\% while the network capacity grew only by
15\%~\cite{gellings}. Apparently as a result of this widening gap between demand and supply, the system became metastable, which
only became apparent when a large part of the power grid broke down.

In the following, we start from an active network and determine under which conditions a collapse to the inactive ground state is likely to
occur.  After an active period, avalanches to the inactive state can only take place if the system initially resides in a metastable active state and then decays to the energetically more favourable inactive state. In the following these requirements are converted into conditions in terms of the constants of the distribution model.

There are indeed two basic restrictions in order to see avalanche effects. The first one concerns the metastability of the active state. The active state should remain intact for a sufficiently long time. Focussing on single-spin-flip dynamics, this requirement corresponds to the impossibility that at zero temperature a spin spontaneously flips from $+1$ to $-1$ while its surrounding remains active (+1). In such a hypothetical process, only the local energy associated with node $i$ changes by an amount $\Delta E^{sf}_i$ given by
\begin{equation}
\Delta E^{sf}_i = 2( H_i + I_i (\sigma_{av} = +1)) .
\end{equation}
At zero temperature, single-spin flips are excluded provided $\Delta E^{sf}_i>0$. The definition of the fields, \eqref{totaalveld}, then leads to the requirement
\begin{equation}
 b < \frac{\avq}{\averageq{1-\mu}} \left ( \frac{1-a}{q_i^{\mu}} + \frac{2-a}{q_i} \sum_{j=1}^{q_i}\frac{1}{q_j^\mu} \right ),  \;\;\forall i \label{2belgen}
\end{equation}
where the sum is over the $q_i$ neighbours of node $i$. The condition depends on the specific local structure around each node in the network.

In order to obtain a simple and useful analytical approximation to this, we can average over the degree of the nearest neighbours. The averaging procedure defined in \eqref{averagen} entails the replacement
\begin{equation}\label{replacement}
\sum_{j=1}^{q_i}\frac{1}{q_j^\mu} \rightarrow \frac{\averageq{1-\mu}}{\avq}q_i.
\end{equation}
Note that the substitution defined in \eqref{replacement} is only exact for $\mu = 0$. For $\mu \neq 0$ it is a topological mean-field approximation. Within this approximation the active state is metastable (i.e., stable against single-spin flips at $T=0$) if for all possible degrees $q_i$,
\begin{equation}
b < \frac{\avq}{\averageq{1-\mu}}\frac{(1-a)}{q_i^{\mu}} + 2-a, \; \forall i \label{mayo}.
\end{equation}
This condition is equivalent to
\begin{equation}
  \ll H_i + I_i (\sigma_{av} = +1)\gg_i  \; >0, \; \forall i,
\end{equation}
which is the metastability criterion for the mean fields defined in \eqref{velden}. The mean-field approximation can therefore by formulated directly in terms of the mean fields.

\eqref{mayo} clearly sets an upper limit to the global demand. If it is too large, a metastable active state is not possible and there will be an ``immediate" decay to the inactive ground state. Depending on the signs of $1-a$ and $\mu$,  two situations can be distinguished. For $\mu>0$ and $a<1$, it suffices that the largest possible degree, $K$, satisfies \eqref{mayo}. On the other hand, for  $\mu>0$ and $a>1$, the smallest possible degree, $m$, is the important one.  For $\mu < 0$, the converse is true.

If more than one spin is allowed to flip at the same time, the criterion \eqref{mayo} must be extended. The energy difference associated with a multiple-spin-flip process can be smaller in magnitude than the energy difference pertaining to a single-spin-flip process. For instance, for a process in which two nearest neighbours  $i$ and $j$ flip, the total energy difference of the double-spin-flip process $\Delta E^{df}_{ij}$ would be
\begin{equation}
\Delta E^{df}_{ij}  = \Delta E^{sf}_i + \Delta E^{sf}_j - 2J(q_iq_j)^{-\mu},
\end{equation}
 which (for $J>0$) is not only always lower than the sum of the energy differences of the two single-spin-flip processes separately ($ \Delta E^{sf}_i + \Delta E^{sf}_j$), but can even be lower than the energy change involved in a single-spin-flip process. This can be appreciated by considering for example the special case $a= b =1$, for which the average value $ \ll \Delta E^{sf}_i \gg_i$ vanishes. The energy difference may in general be reduced further if more than two neighbouring spins can flip simultaneously. In real distribution networks multiple-spin flips can model multiple suppliers failing at the same time, when, for instance, raw materials are exhausted. However, we will not consider these instances but rather allow for random technical failures besides deliberate decisions arising from interactions, making it unlikely that multiple suppliers break down exactly simultaneously. Therefore, for simplicity, single-spin-flip dynamics is assumed throughout the remainder of our paper.

As a second restriction, we require the inactive state to have a lower total energy at $T=0$ than any other state. This condition ensures that the breakdown occurs towards the inactive state rather than to another, glassy state, which minimizes the energy in regions of the parameter space where the local random fields dominate the ferromagnetic interaction. The reason for this limitation is that we wish to study principally the blackout phenomenon in which, after some time, a state with {\em very low activity} is reached.

A sufficient condition for the inactive state (all spins down) to be the ground state (at $T=0$) is that all local quenched random fields have the same sign, in this case negative. The exact requirement is $H_i < 0, \; \forall i$, or
\begin{equation}
b >  (1-a) \frac{\avq}{\averageq{1-\mu}} \left ( \frac{1}{q_i^{\mu}} + \frac{1}{q_i} \sum_{j=1}^{q_i}\frac{1}{q_j^\mu} \right ),  \;\;\forall i
\label{2fransen}
\end{equation}
In the mean-field approximation adopted in the previous paragraph this becomes
\begin{equation}
b > \frac{\avq}{\averageq{1-\mu}}\frac{(1-a)}{q_i^{\mu}} + 1-a, \; \;\forall i \label{wontonton}.
\end{equation}
\eqref{2fransen} and its simplified version \eqref{wontonton} provide a threshold for the demand sufficient to observe decays to the inactive state. When the demand is  smaller than this threshold, the absolute stability of the inactive state is not guaranteed. Then the system may remain stable in the active configuration (all spins up) or evolve to a glassy state. Once more, different regimes can be identified according to the values of $a$ and $\mu$. For example, for positive $\mu$, if $a<1$ it suffices that the smallest possible degree satisfies \eqref{wontonton}. However, if $a>1$, the relevant quantity is the largest degree. These statements are to be interchanged if $\mu$ is negative.

\begin{figure}[t]
 {\includegraphics[width = \textwidth]{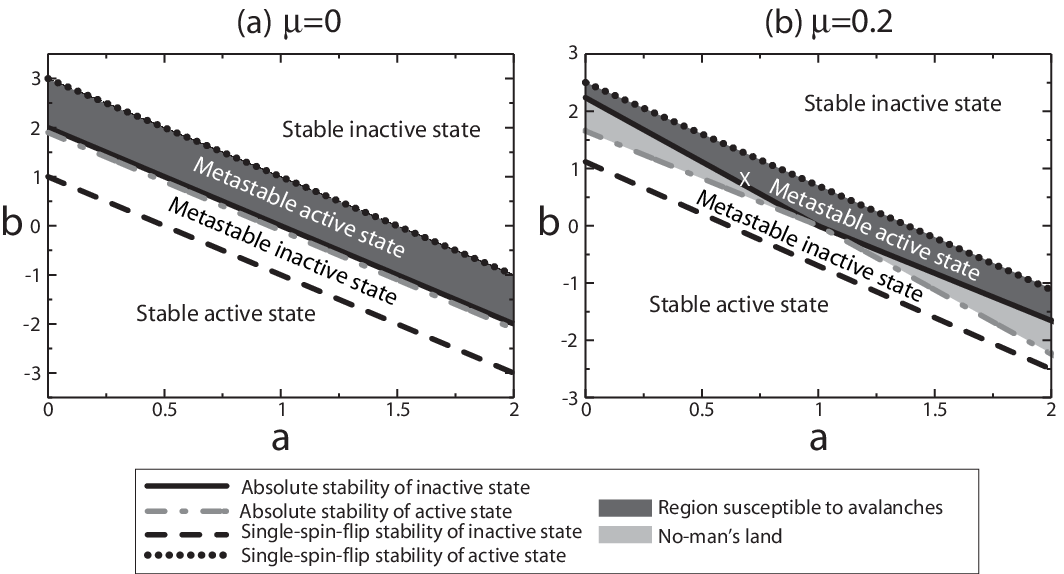}}
 \caption{Phase diagram in $(a,b)$-space of the distribution model for a scale-free network with 1000 nodes, topological exponent $\gamma = 3$ and minimal degree $m=2$. Part (a) corresponds to $\mu = 0$, while part (b) pertains to $\mu = 0.2$. The uppermost (dotted) lines indicate the upper boundaries of metastable active states for single-spin-flip dynamics at $T=0$. The second uppermost lines (solid) denote the absolute stability limit of the inactive state at $T=0$. The dark-grey region filling the space between the uppermost and second uppermost lines corresponds to a regime in which, at $T>0$, a single-spin flip can cause an avalanche starting from a metastable active state. The next lower line (dash-dotted; coincident with the foregoing one in part (a) but not in part (b) of the figure) marks the absolute stability limit of the active state at $T=0$. In Fig.~(b), the ground state is unknown in the no-man's land between the two absolute stability lines, indicated by the light-grey shading.
 Finally the lowermost lines delineates the lower boundaries of metastable inactive states for single-spin-flip dynamics at $T=0$.  In part (b) the white cross indicates the specific values of $a$ and $b$ used in Fig. 3(a).
 \label{fasediagramvsmu}}
 \end{figure}

The two conditions \eqref{2belgen} and \eqref{2fransen}, or their mean-field approximations \eqref{mayo} and \eqref{wontonton}, determine the region in a $(a,b)$-phase diagram where collapses of an active network to an inactive one should be observable. A graphical representation of the phase diagram for distribution models with $\mu =0$ and $\mu =0.2$ can be found in Fig.~\ref{fasediagramvsmu}. The shaded region on the Figure indicates the ranges of $a$ and $b$ within which the two requirements are satisfied for a scale-free network with topological constant $\gamma = 3$, minimal degree $m = 2$ and 1000 nodes. The uppermost line (dotted) is the numerically exact upper bound for the stability of the active state against single-spin flips at $T=0$, the so-called metastability limit, given by \eqref{2belgen}. This exact condition was determined by simulation of the model on scale-free networks.

 The networks were generated in two steps using the uncorrelated configuration model. In a first step, $N$ nodes are created, each with their own degree, chosen randomly according to the distribution $P(q)$, which is a decreasing power law. The minimal degree is $m=2$, while the maximal degree is set to $K = \sqrt{N}$. For $\gamma > 3$ this is not a restriction, since the maximal degree satisfies $K \propto N^{1/(\gamma - 1)}$ \cite{cohen}. On the other hand, for $\gamma < 3$ this restriction avoids correlations in the network \cite{catanzaro}. In a second step, links are laid randomly between all the nodes, with the constraint that at the end of the linking procedure every node should have the degree it was given in the first step. Self linking and multiple linking are avoided. The results obtained from different network realizations differ slightly. In practice, averaging over 10 networks is sufficient to obtain accurate reproducible results. In this manner we determine the values of  $a$ and $b$ for which \eqref{2belgen} and \eqref{2fransen} are satisfied for all nodes. The latter condition leads to the second uppermost line (solid) in the figures (straight line for $\mu=0$; line consisting of two straight segments, with a break at $a=1$, for $\mu \neq 0$). Above this line the inactive state is, with certainty, the ground state (at $T=0$). The region in between the two lines discussed so far is susceptible of avalanches of spin flips, i.e., blackouts of the network, and is shown shaded (in dark grey).

 The model possesses an overall symmetry, which is reflected in the phase diagrams in Fig.1. Inspecting \eqref{hamiltoniaan} and \eqref{totaalveld}, we conclude that the full Hamiltonian is invariant under the transformation: $a\rightarrow a' = 2-a$; $b \rightarrow b' = - b$; $\sigma \rightarrow \sigma' = - \sigma$ (flipping all the spins). This symmetry implies that the lines in the phase diagram pertaining to the active state (all spins up) can easily be drawn by applying the above transformation of $b$ and $a$ on the lines associated with properties of the inactive state (all spins down). In particular, the lowermost line (dashed) marks the metastability limit of the inactive state, and the second lowermost line (dash-dotted;  broken for $\mu \neq 0$) indicates the limit of absolute stability of the active state. Below this line the active state is, with certainty, the ground state (at $T=0$). A remarkable feature of the phase diagram now emerges. While for $\mu = 0$ the two limits of absolute stability coincide, so that there are only (two) ferromagnetic ground states, a new zone appears for $\mu \neq 0$. In this zone, which we term ``no-man's land", the ground state is not with certainty ferromagnetic. It may be a glassy state, characterized by local random fields of either sign (up or down) that try to orient the spins along their direction, at low $T$, as they compete with the ferromagnetic spin-spin coupling. The no-man's land is shaded in light grey. Note that the width of this no-man's land vanishes for $b = 0$, and also, for $a = 1$, for which the random fields are trivially all of the same sign, determined by the value of the remaining free parameter $a$ or $b$, respectively.

In general, the phase diagram depends on the network topology. However, the case $\mu = 0$ provides an exception. As mentioned before, the single parameter $a_0 = a + b/2$ suffices for describing the supply and demand functions, if $\mu$ vanishes. The conditions \eqref{mayo} and \eqref{wontonton}, which are {\em exact} for $\mu = 0$, then correspond to the simple conditions $1 < a_0 < 3/2$.  Note that these can also be obtained directly starting from \eqref{richtingkust}. These conditions imply that there are no distinctions between different networks at zero temperature if $\mu$ vanishes. Indeed, for $\mu = 0$, the sign of the quenched random field is independent of the network structure, as can be seen in \eqref{richtingkust}. Therefore, the topology of the network does not affect the state of the system at $T=0$. This prediction is verified by comparing simulations on scale-free networks with simulations on random  Erd\H{o}s-R\'{e}nyi graphs. In an  Erd\H{o}s-R\'{e}nyi network all pairs of edges have  equal probability to be linked, which implies a Poissonian degree distribution. We verified that
  the condition \eqref{2belgen} leads to one and the same straight line in the $(a,b)$-plane on an Erd\H{o}s-R\'{e}nyi network and on a scale-free network with $\gamma = 3$. The same is true for \eqref{2fransen}. However, if $\mu \neq 0$, the network topology exerts a crucial influence on the phase diagram. We verified that, for example, both with respect to  the mean-field conditions and the exact ones, the phase diagram is different for networks with different values of $\gamma$ \cite{thesisSVL}.

In the following we go into more detail as regards the interesting comparison of the mean-field approximation and the (numerically) exact results for the boundaries in the phase diagram. As we already noted, the topological mean-field approximation is (only) exact for $\mu = 0$. The simulations confirm this. To estimate the quantitative difference between, for example \eqref{2belgen} and \eqref{mayo} we calculate the variance, over all network realizations, of a quantity associated with the sum in the last term of \eqref{2belgen}. We obtain
\begin{equation}
{\rm Var}\left ( \frac{1}{q_i} \sum_{j=1}^{q_i} q_j^{-\mu} \right )  =  \frac{1}{q_i} \left ( \frac{\averageq{1-2\mu}}{\avq} - \frac{\averageq{1-\mu}^2}{\avq^2}\right )
\label{varian}
\end{equation}
Consequently, the ratio of the standard deviation to the mean is given by
\begin{equation}
{\rm SD}\left ( \frac{1}{q_i} \sum_{j=1}^{q_i} q_j^{-\mu}\right )/   \ll \frac{1}{q_i} \sum_{j=1}^{q_i} q_j^{-\mu} \gg_i  \; =  \frac{1}{\sqrt{q_i}} \sqrt{ \frac{\averageq{1-2\mu}\avq}{\averageq{1-\mu}^2}-1}
\label{standdev}
\end{equation}
This result allows us to estimate the importance of the {\em random field fluctuations}. Clearly, the hubs are least affected by the topological disorder since the SD scales with their degree $q_i$ as $1/\sqrt{q_i}$. The amplitude of this power law depends on $\mu$ in a manner which is easy to interpret. For example, for $\mu = -1, 0.5$ or $1$ the variance is proportional to $ \avq \averageq{3} - \averageq{2}^2$, $ \avq  - \averageq{1/2}^2$ or $\averageq{-1} - \avq^{-1}$, respectively.

We now proceed to assess quantitatively the difference between the exact {\em metastability limit} \eqref{2belgen} and the approximate one \eqref{mayo} for $\mu = 0.2$. Clearly, as Fig. 2(a) shows, the mean-field approximation leads to a continuous piece-wise linear curve which is broken at $a =1$, since for $a < 1 (>1)$ the right-hand-side of \eqref{mayo} is minimized by the maximal (minimal) degree present in the network. Interestingly, the exact (dotted) curve for $\mu=0.2$, shown in Fig. 2(a) and also in Fig. 1(b), is a straight line, without any singularity, since the second term in the right-hand-side of \eqref{2belgen} is typically minimized by a node $i$ of degree $q_i=2$ connecting two hubs ($q_j \gg 1$). This minimal value lies some 3 to 4 standard deviations below the mean. Consequently, the entire right-hand-side of \eqref{2belgen} is minimized by one and the same node $i$ with a low value of $q_i$ for all $a \in [0,2]$. For $\mu$ larger than some threshold this is no longer the case and the numerically exact metastability limit is no longer a straight line but a gently bent concave curve. Still, its shape appears smooth and is therefore qualitatively different from the broken curve found in the mean-field approximation. This is conspicuous in Fig. 2(b) where both are shown for $\mu = 1$.

\begin{figure}[t]
\includegraphics[width = \textwidth]{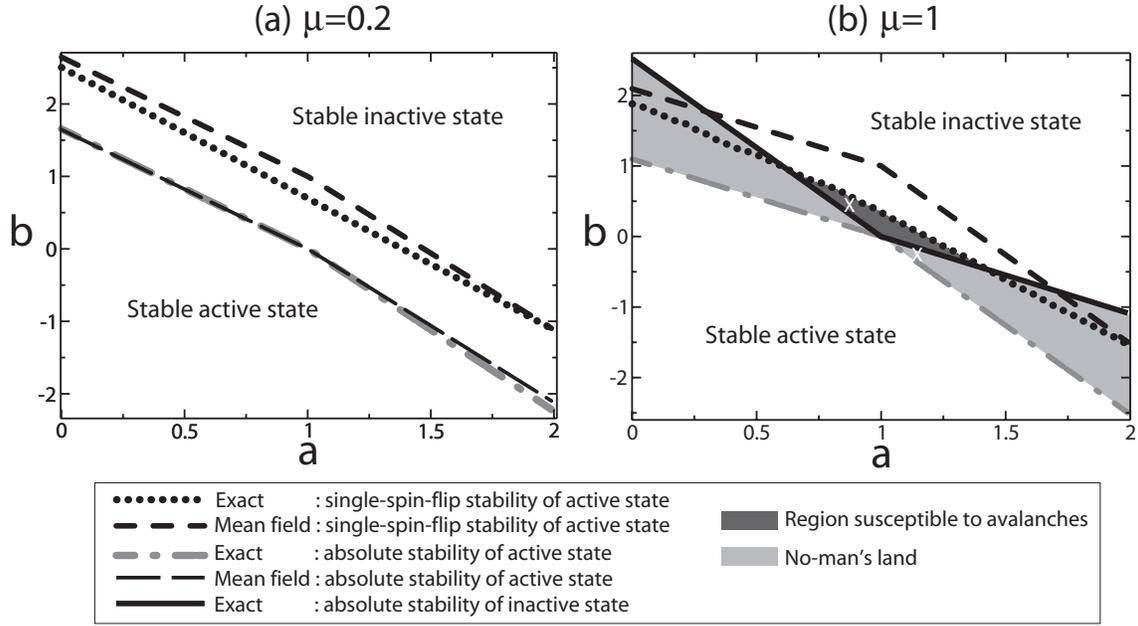}
 \caption{(a) Comparison of two mean-field and two exact phase boundaries for a distribution model with topological exponent $\gamma =3$ and delivery exponent $\mu = 0.2$, with 1000 nodes and $m = 2$. Shown are the upper limit of the metastable active state for single-spin-flip dynamics at $T=0$ (uppermost curves; dotted line for the exact result and dashed line for the mean-field approximation) and the upper limit of absolute stability of the active state at $T=0$ (lowermost curves; dash-dotted line for the exact result and dashed line for the mean-field approximation);  (b) Phase diagram for $\gamma = 3$ and $\mu = 1$. The uppermost broken curve is the upper limit of metastability of the active state for single-spin flips at $T=0$ in the mean-field approximation (dashed line). The smooth curve below it (dotted) is the numerically exact result for that same metastability limit. Also shown are two crossing lines. The upper segments (solid) form the lower limit of absolute stability of the  inactive state at $T=0$, while the lower segments (dash-dotted) mark the upper limit of absolute stability of the active state at $T=0$. The region that is (most likely) susceptible of avalanches is the small triangle shaded in dark grey. The no-man's land, with an unknown ground state, is shaded in light grey. The white crosses indicate the specific values of $a$ and $b$ used in Figs. 3(b-d).
 \label{fasediagramvsmu2}}
 \end{figure}

We repeat our analysis for the {\em absolute stability limit} of the active state. This is the curve below which the active state is with certainty the ground state (at $T=0$). It is derived numerically exactly by applying the transformation $a \rightarrow a' = 2-a$; $b \rightarrow b' = - b$ to \eqref{2fransen} and in the mean-field approximation by applying the same symmetry to \eqref{wontonton}, which in both cases simply reverses the inequalities concerned. In contrast with the metastability limit, the slope of the absolute stability limit displays a discontinuity at $a =1$ for both the exact and the mean-field versions, as can be seen from the fact that both \eqref{2fransen} and \eqref{wontonton} contain the prefactor $1-a$. For $a < 1$ the limit is defined through a hub (typically the node with the highest degree) and for $a > 1$ through a node with a low degree. For the former case, the mean-field approximation is accurate as expected since for hubs random field fluctuations are small, being proportional to $1/\sqrt{q_i}$. For the latter case, random field fluctuations are more important and, indeed, a clear difference emerges between the mean-field upper bound and the exact one. In Fig.~\ref{fasediagramvsmu2}(a) we compare the exact and the mean-field curves for $\mu = 0.2$, both for the metastability limit of the active state and for the absolute stability limit of the same state.

The region in which avalanches can occur depends sensitively on the exponent $\mu$. Comparing Fig.~\ref{fasediagramvsmu}(a), Fig.~\ref{fasediagramvsmu}(b) and Fig.~\ref{fasediagramvsmu2}(b), the size of the region in which blackouts are possible shrinks as $\mu$ increases. The same trend is observed when $\mu$ is decreased from zero \cite{thesisSVL}. We conclude that the region in parameter space in which our two criteria for avalanches are fulfilled shrinks as $|\mu|$ increases. This region is also sensitive to the topological exponent $\gamma$. For instance, for $\mu = 1$, when $\gamma$ is decreased from 5 to 2 there appear typically more nodes with higher degrees. Consequently, the mean degree $\avq$ increases. This raises the absolute stability limit of the inactive network for $a < 1$ and lowers the metastability limit of the active state for $a > 1$, while leaving the other segments of the phase boundaries unchanged (as can be seen qualitatively by inspecting the equations for the mean-field (meta-)stability criteria). As a consequence the region susceptible of avalanches is squeezed more tight about the center ($a = 1, b = 0$) of the phase diagram. At this point we would like to mention other studies in which the resistance against cascades was studied as a function of geometrical disorder. In models based on critical loads and percolation in interdependent networks it was found that a heterogeneous network is less resistant against higher loads than a homogeneous one \cite{motter,crucitti,buldyrev}. In our model degree heterogeneity (low $\gamma$) appears to strengthen the network at low $a$ but to weaken it at high $a$.

Also the size of the network influences the region available for collapses, mainly through the maximal degree which depends on $N$. The region in which avalanches are observed decreases if more suppliers are present in the network, as can be seen from Eqs.~(\ref{mayo}) and (\ref{wontonton}). However, a  numerical evaluation of these criteria indicates that the effect of the number of suppliers is rather small compared to the effects of the values of $\mu$ and $\gamma$ \cite{thesisSVL}.
We conclude that the ranges of the demand parameters $a$ and $b$ for which collapses occur are influenced most strongly by the network topology and the degree dependence of the delivery in the distribution system.

\section{Collapse properties}
\subsection{Distribution model at finite temperature}
In real-life distribution systems, individual suppliers can fail to deliver their goods, for instance due to a defect or malfunction in the production process. After repair the delivery resumes. Therefore, we introduce a generic ``temperature" $T$ in the distribution model to quantify the rate of random spin fluctuations, from the active to the inactive state and back.  At sufficiently high temperatures, the mean spin magnetization $M = \sum_i \sigma_i /N$, which represents the mean network activity, tends to zero as a function of $T$ according to a Curie-Weiss-type law, i.e., $M \propto 1/T$ for $T \rightarrow \infty$ (cf. a paramagnet in a small external field). We will, however, focus mainly on lower temperatures, still in the ferromagnetic regime, for which a network failure can occur.

\begin{figure}[t]
\subfigure[$\mu = 0.2, a = 0.7, b = 0.9, k_BT/J = 0.08$\label{avalanche1}]{\includegraphics[angle = 270,width = 0.5\textwidth]{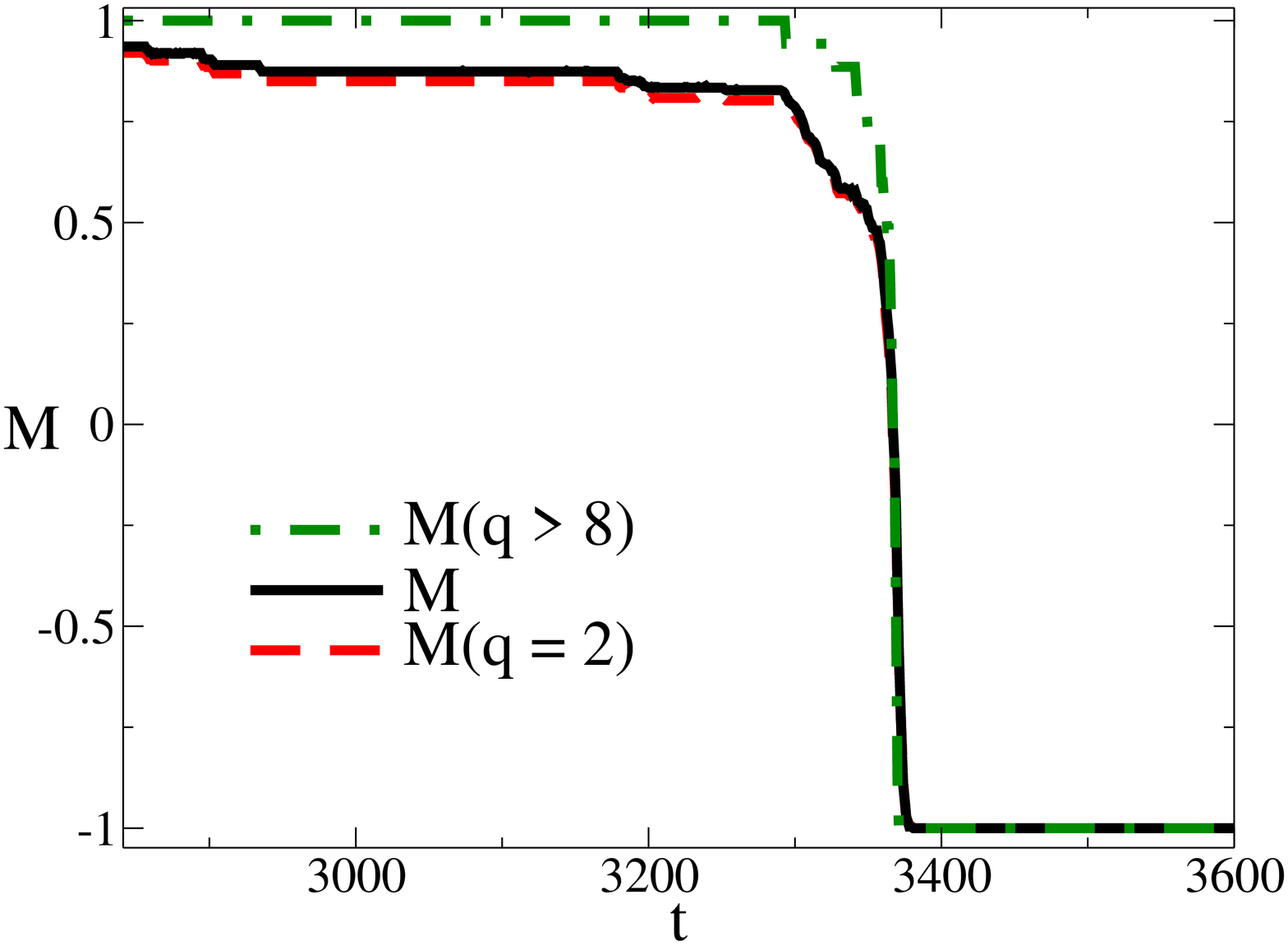}}
\subfigure[$\mu = 1,a = 1.15, b = -0.25, k_BT/J = 0.03$\label{avalanche4}]{\includegraphics[angle = 270,width = 0.5\textwidth]{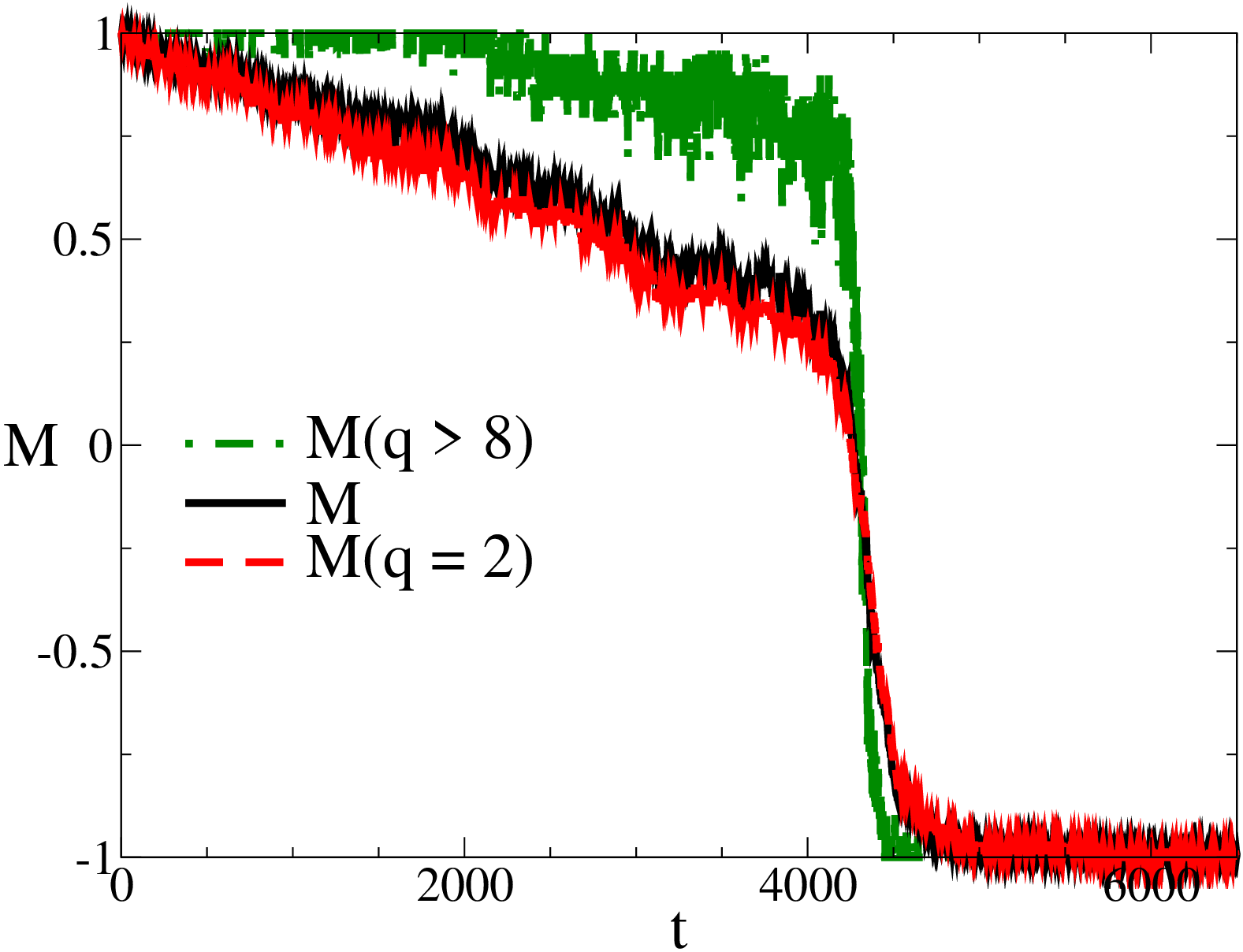}}\\
\subfigure[$\mu = 1, a = 0.85, b = 0.4, k_BT/J = 0.028$\label{avalanche3}]{\includegraphics[angle = 270,width = 0.5\textwidth]{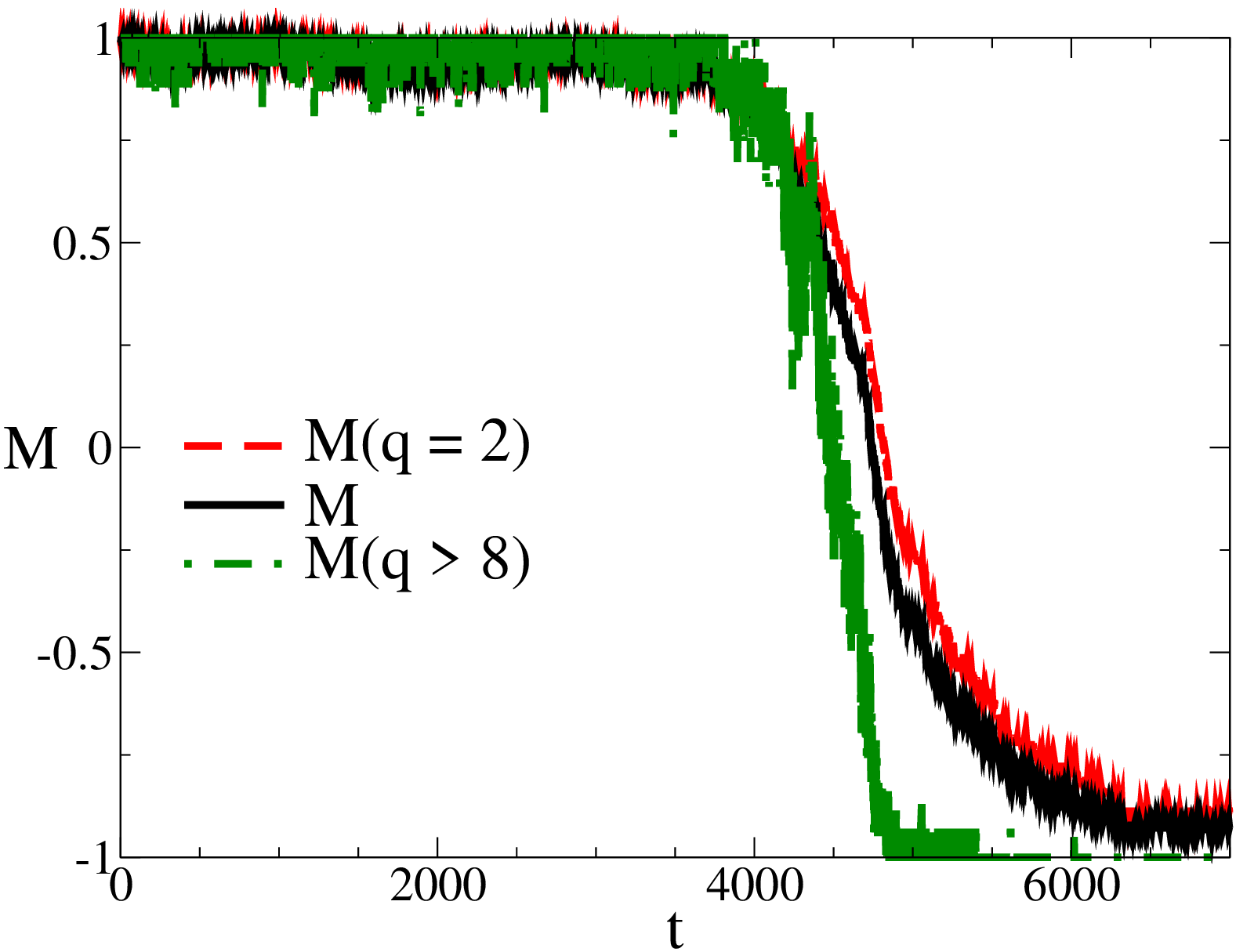}}
\subfigure[$\mu = 1, a = 0.85, b = 0.4, k_BT/J = 0.03$\label{avalanche2}]{\includegraphics[angle = 270,width = 0.5\textwidth]{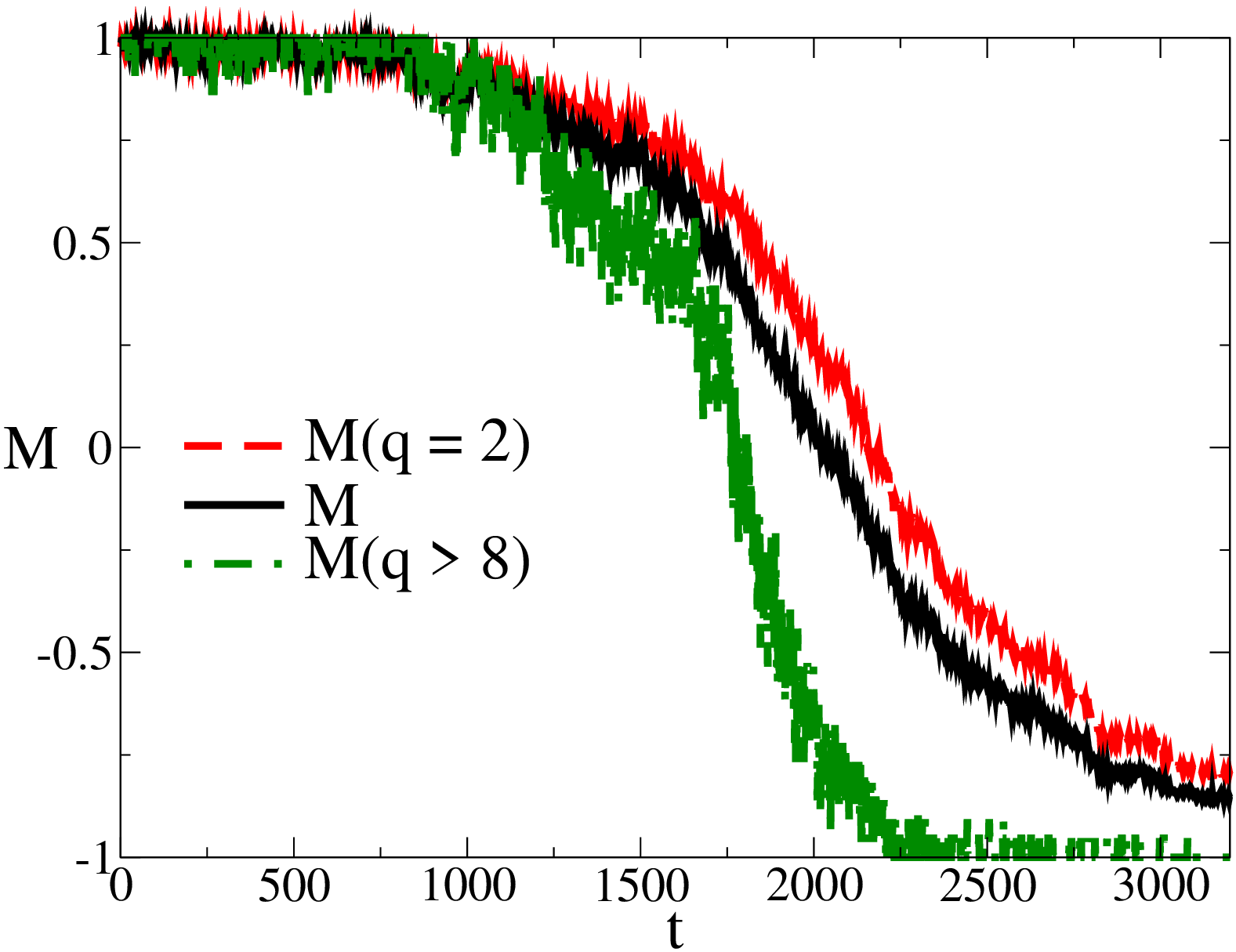}}
 \caption{(Color online) Typical examples of breakdowns in distribution networks. The graphs show the mean magnetization $M$ in the network as a function of time. All figures are for the distribution model on scale-free networks with $\gamma = 3$ and 1000 nodes.   The values of the constants $a$, $b$ and $\mu$ differ in the different subfigures. They correspond to the positions of the crosses in the phase diagrams Fig. 1(b) and Fig. 2(b) for $\mu = 0.2$ and $\mu = 1$, respectively. The solid (black; middle) curve indicates the mean magnetization, averaged over all possible node degrees. The dot-dashed (green; top in (a) and (b), bottom in (c) and (d)) curve gives the mean magnetization of the nodes with 8 or more neighbours ($M(q>8)$) while the dashed (red; bottom in (a) and (b), top in (c) and (d)) curve shows the mean magnetization of the nodes with 2 neighbours ($M(q=2)$).
 \label{collapses}}
 \end{figure}

We perform simulations on scale-free graphs which are constructed using the uncorrelated configuration model introduced in the previous Section. All networks contain 1000 nodes, the minimal degree of which is $m=2$ and the maximal one $K = \sqrt{1000} \approx 32$. Spins are updated using single-spin-flip dynamics with the Metropolis updating rule.
The simulations start from a metastable active state. Collapses can be found for various values of the parameters  $\mu$, $a$, $b$ of the distribution model and for different values of the topological exponent $\gamma$. Some examples are shown in Fig.~\ref{collapses}, in which the time evolution of the magnetization is plotted. A single time step corresponds to 1000 single-spin flips.
Different types of collapses are observed in different ranges of model parameters, as we will illustrate in the following.

A first point of attention concerns the nodes which initiate the collapse. Apart from the mean magnetization of the network, Fig.~\ref{collapses} also shows the mean magnetization of the highly connected nodes (with at least 8 links) and that of the poorly connected nodes (with only two neighbours). In the first and second collapses, Fig.~\ref{avalanche1} and Fig.~\ref{avalanche4}, the poorly connected nodes exhibit the largest fluctuations before the collapse. The nodes with a low degree thus initiate the breakdown. However, the actual collapse only takes place when also the hubs start to flip. As long as the hubs remain active, their large influence in the network prevents a blackout. The opposite behavior is found in the avalanches shown in Fig.~\ref{avalanche3} and Fig.~\ref{avalanche2}, in which the hubs initiate the collapse. Again, the mean magnetization remains positive until also the multitude of less-connected nodes starts to collapse. Thus in all four cases only a certain subset of nodes initiates the network collapse, but the network undergoes a full breakdown as a consequence of a collective effect, in which all nodes are involved. Our model thus displays the cooperative character of distribution networks as described in the Motivation.

A simple mean-field argument suggests a condition for determining whether the hubs or the poorly connected nodes initiate the collapse. The collapse is initiated by the nodes which have the largest fluctuations in the active state. We introduce an {\em activation temperature} for a node with degree $q$, $T_{act}(q)$, by equating its thermal fluctuation energy to the average energy needed to spin flip a node with degree $q$ from $+1$ to $-1$. When $T > T_{act}(q)$ the spin of the node will undergo significant thermal fluctuations. Within the mean-field approximation, $T_{act}(q)$ is given by
\begin{equation}\label{crit_temp}
 k_B T_{act}(q_i)=2\ll H_i+I_i(\{\sigma\})\gg_i
=Jq_i^{1-2\mu}(1-a)+Jq_i^{1-\mu}\frac{\langle
q^{1-\mu}\rangle}{\langle q\rangle}\left(1-
a-b+\sigma_{av}\right).
\end{equation}
We focus on the degree-dependent activation temperature when the network is in an (almost) fully active state, thus with $\sigma_{av} \lesssim 1$.  Then, the nodes with the smallest  $T_{act}(q)$ display the largest fluctuations in the active state and can thus initiate the collapse of the network when $T$ is slowly raised from zero. Note that whether the activation temperature is either increasing or decreasing as a function of the node degree, depends on the signs of $1-a$, $\mu - 1/2$ and $2-a-b$ (assuming $\sigma_{av}=1$). Let's test these ideas against some of the simulations of network breakdowns shown in Fig.~\ref{collapses}. For example, using the model parameters associated with Fig.~\ref{avalanche1}, we obtain, with $\sigma_{av} \lesssim 1$, $k_BT_{act}(q)\approx J(0.30 \,q^{0.6}+0.31\, q^{0.8})$, which implies that the activation temperature increases with the node degree. Therefore, the poorly connected nodes should initiate the collapse, which is confirmed by the simulations. Taking the parameters associated with Fig.~\ref{avalanche2}, we find $k_BT_{act}(q)\approx J(0.28 + 0.15\, q^{-1})$, so that the highly connected nodes should show the largest fluctuations in the active state. The simulations confirm that the hubs indeed initiate the network collapse.

Information about the type of nodes that initiates a collapse is of great interest in real-life networks. If hubs tend to be the most fragile and thus most fluctuating suppliers, it is best to invest more effort in protecting them rather than the least connected nodes. Of course the converse is true if the poorly-connected nodes are more prone to initiate the collapse. The criterion of \eqref{crit_temp} could therefore be used to strengthen networks purposefully against the consequences of accidental malfunction of distribution centers.

\subsection{Effective strength of thermal fluctuations}
As can be seen in Fig.~\ref{collapses}, the magnitude of the thermal fluctuations depends not only on the value of $T$ but also on that of other parameters of the distribution model. Even if the demand is fixed (constant $a$ and $b$), thermal fluctuations may still be reduced or amplified  depending on the network structure, through the topological exponent $\gamma$, and depending on the delivery in the model, through the delivery exponent $\mu$.
The starting point of our further analysis is the mean Hamiltonian, Eq.~(\ref{meanham}). The expression for this Hamiltonian together with the  expressions for the amplitude of the fluctuations of the non-local term in the quenched random fields, Eqs.~(\ref{varian}) and (\ref{standdev}), suggest that, subject to conditions to be specified, the dependence on the network topology and on $\mu$ might be captured by the single parameter $\averageq{1-\mu}^2/\avq$. This prompts us to test the conjecture that the mean Hamiltonian may possess the following scaling property,
\begin{equation}
  \langle\mathcal{H}(\sigma_{av})\rangle \approx -NJ\sigma_{av}\frac{\averageq{1-\mu}^2}{2\avq}f(a,b,\sigma_{av}),\label{australie}
\end{equation}
where the function $f(a,b,\sigma_{av})$ is independent of $\mu$ and $\gamma$. This Ansatz is most likely to be valid in at least one of the two following circumstances. Firstly, if $a$ is sufficiently close to one ($a \approx 1$), the first term in \eqref{meanham} is negligible compared to the second one and \eqref{australie} holds with $f(a,b,\sigma_{av}) \approx -b+\sigma_{av}$. The second situation occurs for large networks and small enough $\mu$, i.e., $|\mu| \ll \gamma -2$, which leads to $\averageq{1-2\mu}\approx\averageq{1-\mu}^2/ \avq$, as can readily be shown analytically. Indeed, in the thermodynamic limit, $N, K \rightarrow \infty$, and converting sums over the degree distribution to integrals \cite{opm}, one finds
\begin{equation}
\frac{\averageq{1-2\mu}\avq}{ \averageq{1-\mu}^2} \approx 1 + \frac{\mu^2}{(\gamma-2)^2}\frac{1}{1+\frac{2\mu}{\gamma -2}}
\end{equation}
Note that, in view of \eqref{standdev}, the condition $|\mu| \ll \gamma -2$ thus ensures that the random-field fluctuations are small. Under these circumstances \eqref{meanham} indeed takes the scaling form of \eqref{australie} with $f(a,b,\sigma_{av}) = 2 - 2a-b+\sigma_{av}$. Although, strictly speaking, this last result is only valid for the range of $\mu$ specified above, numerical inspection shows that it is a good approximation even for   values of $\mu$  up till about unity, provided the network is large enough for the $K$-dependence of the averages to be negligible. In general, the approximation remains useful also for finite networks, as numerical analysis shows. In the remainder of this section we assume $\mu \geq 0$.

According to Boltzmann statistics and using the Ansatz of \eqref{australie}, the probability to observe a network with mean spin $\sigma_{av}$ satisfies
\begin{equation}
\mathcal{P}(\sigma_{av}) \propto \exp\left(\frac{NJ\averageq{1-\mu}^2}{ k_B T\avq}\frac{\sigma_{av}f(a,b,\sigma_{av})}{2}\right).\label{boltzmann}
\end{equation}
Apart from the constants related to the demand function, all model parameters ($\mu$, $\gamma$ and $T$) are only present in the first factor in the exponential function. We therefore
absorb the dependence on the delivery exponent, the topology and the temperature into a single parameter $\Theta$, defined as
\begin{equation}\label{theta}
 \Theta = \frac{\avq}{\averageq{1- \mu}^2}T.
\end{equation}
For systems with fixed $a$ and $b$, the finite-temperature behavior of the system is thus controlled by $\Theta$, which acts as an {\em effective temperature}.
In the next subsections, the effects of both the delivery exponent $\mu$ and the topological exponent $\gamma$ on finite-temperature collapses are investigated separately.
\subsubsection{Effect of the delivery exponent $\mu$}

We now focus on a distribution model with fixed demand constants $a$ and $b$ on a scale-free network with fixed topological constant $\gamma$ and investigate the effect of different values of the delivery exponent $\mu \in [0,1]$. Since $\averageq{1-\mu}$ decreases with increasing $\mu$, the effective temperature $\Theta$ of the network defined in \eqref{theta} increases as a function of $\mu$. Thermal spin fluctuations are thus larger in a distribution model with larger $\mu$ than in a model with smaller $\mu$ at the same temperature $T$.  As $\mu$ increases, the network thus becomes more vulnerable, since the collapse temperature decreases. The considered  distribution network could therefore be effectively strengthened by a decrease of  $\mu$, i.e., by rendering the amounts of goods delivered to each customer more homogeneously distributed.

In addition to this ``mean-field" effect there is an effect of the fluctuations of the quenched random nearest-neighbour couplings, or {\em random-bond fluctuations}. These are of topological origin and induced by the (quenched) degree fluctuations of the nodes. To understand this we recall that previous studies have shown that the (equilibrium) critical temperature $T_c$ of the model in zero external field depends rather sensitively on the values of $\gamma$ and $\mu$. In particular, for $\mu = 0$, the critical temperature is finite as long as the second moment of the degree distribution is finite, but diverges as a function of the network size $N$ when $\averageq{2}$ diverges \cite{Aleksiuk,dorogov}, which is the case for $\gamma \leq 3$. For $\mu \neq 0$, whether or not $T_c$ is finite (for an infinite network) is determined by whether or not the effective exponent $\gamma' = (\gamma - \mu)/(1 - \mu)$ exceeds the value 3 \cite{giuraniuc1}. Therefore, it is important to check also the value of $\gamma'$ in our distribution systems. In particular, when $\mu$ is increased (from 0 towards 1), the exponent $\gamma'$ increases and the hubs become less numerous and less pronounced. Consequently, $T_c$ decreases and this also renders the thermal spin fluctuations more important. This effect will be most relevant for networks with $\gamma \leq 3$ and $\gamma' > 3$.

The effects we described are confirmed by simulations in which the mean magnetization is studied versus temperature. In each such simulation we initialize our network in the metastable active state and we update the spins using single-spin flips during 4000 time steps.  At sufficiently low temperatures, the metastable state remains stable during the entire simulation and the final magnetization (after 4000 time steps) remains close to one. Repeating this procedure for a sequence of fixed temperatures, {\em upon increase of the temperature} one observes a transition to the regime in which the active state is no longer metastable on the time scale of the simulations. This (non-equilibrium) transition, which takes place at a {\em breakdown temperature} $T_b$, is marked by a magnetization jump to a negative value. An inactive state, with $\sigma_{av} \gtrsim -1$, is then reached before the end of the simulations. At still higher temperatures, the final magnetization becomes smaller in absolute value and displays a Curie-Weiss behavior reminiscent of the paramagnetic state. Note that, in the absence of symmetry-breaking fields, the final magnetization would approach zero rather sharply at the equilibrium critical temperature $T_c$. We also verified this in our simulations. Obviously, $T_b < T_c$.

\begin{figure}[t]
  \subfigure[]{\includegraphics[angle = 270,width = .5\textwidth]{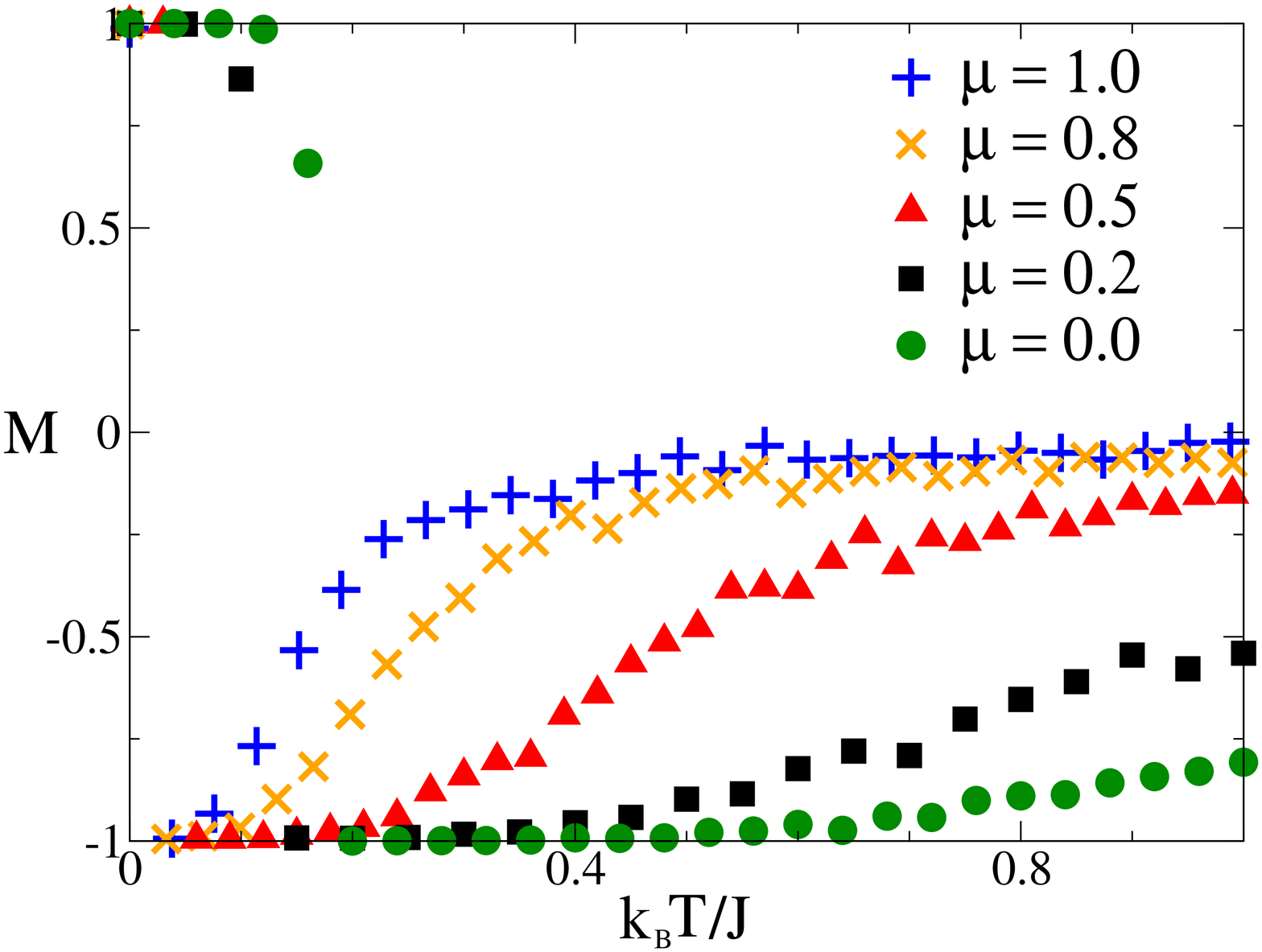}\label{effectmu1}}
  \subfigure[]{\includegraphics[angle = 270,width = .5\textwidth]{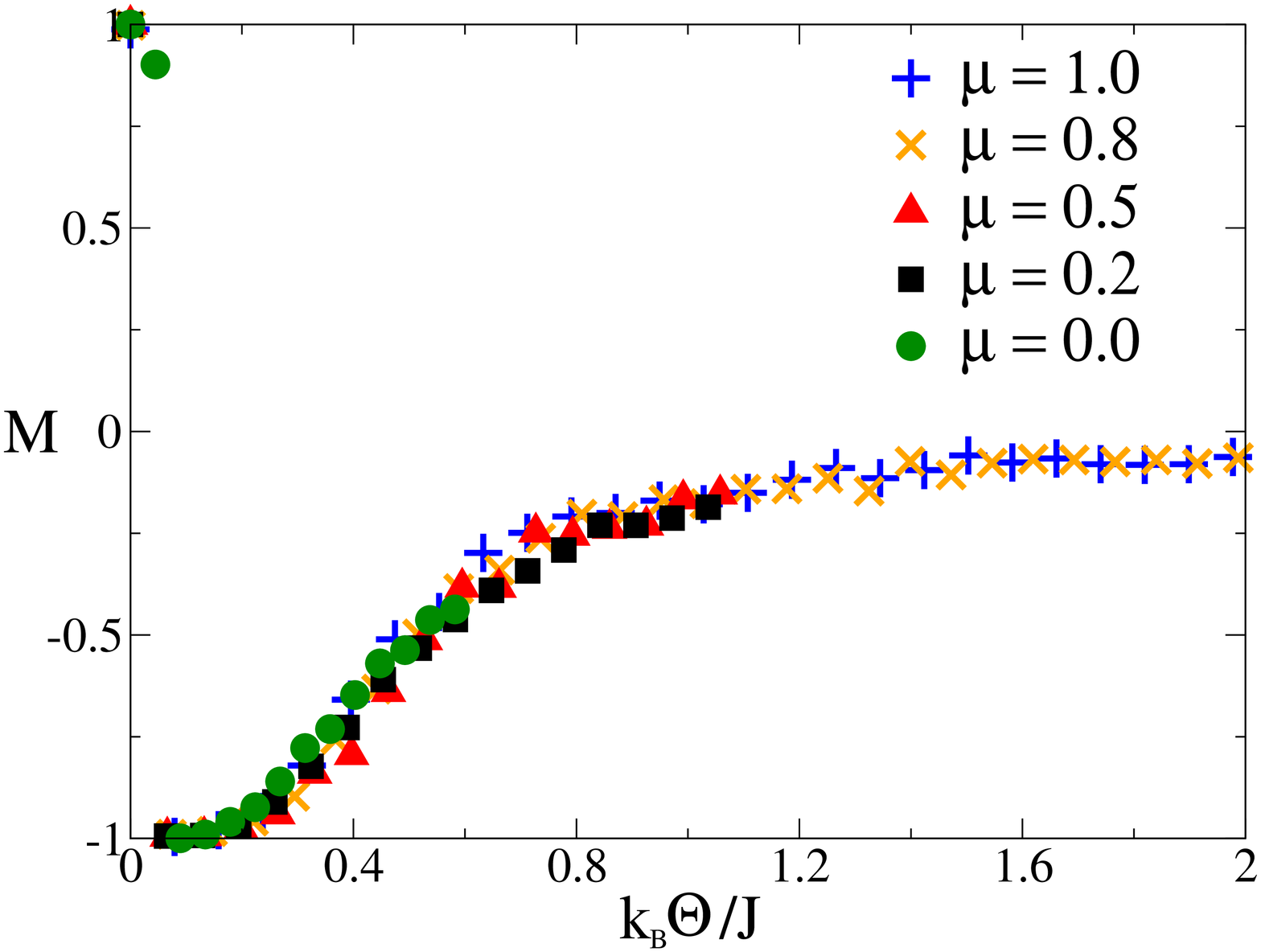}\label{effectmu2}}\\
    \subfigure[]{\includegraphics[angle = 270,width = .5\textwidth]{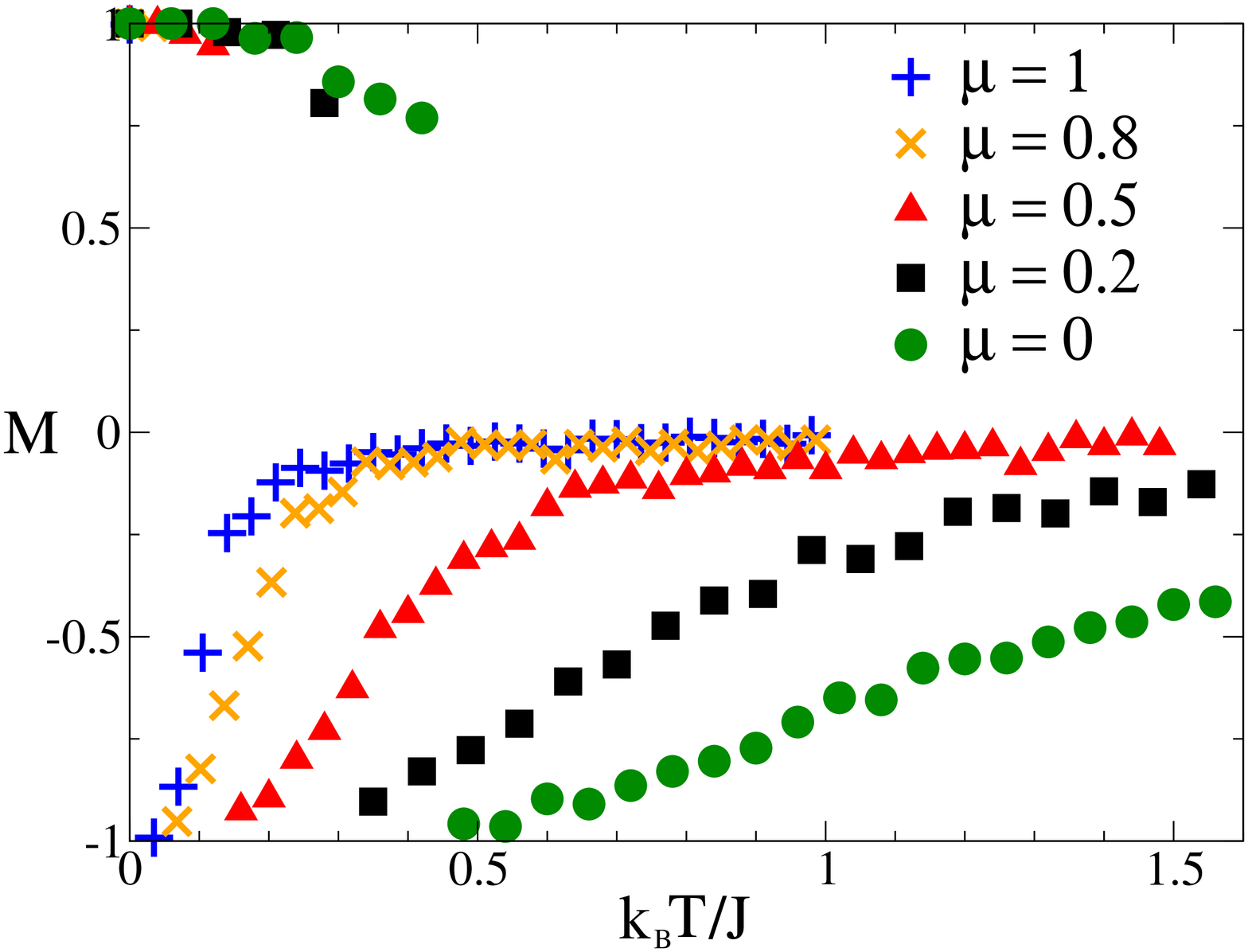}\label{effectmu3}}
  \subfigure[]{\includegraphics[angle = 270,width = .5\textwidth]{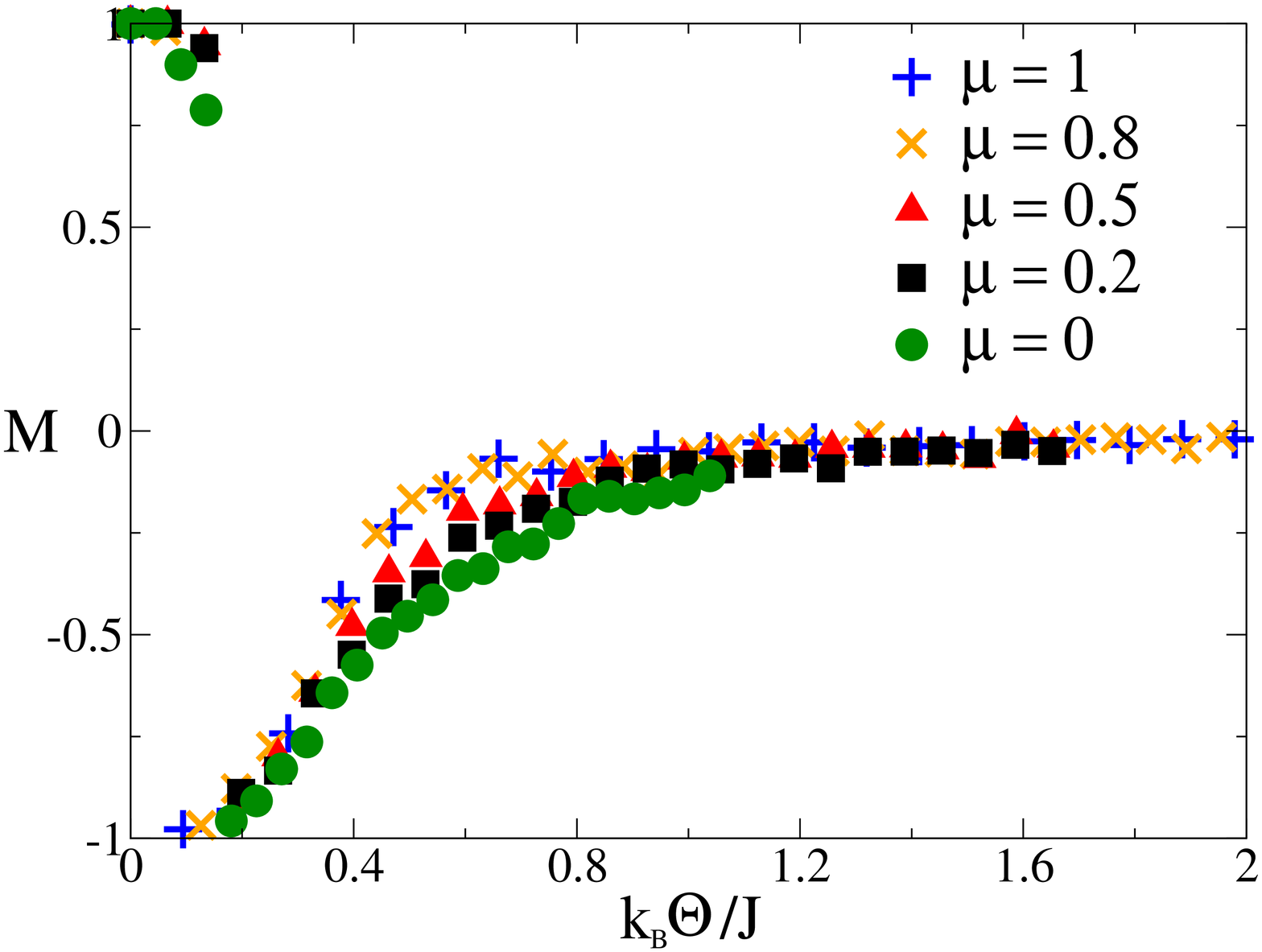}\label{effectmu4}}
\caption{(Color online) Effect of the delivery exponent $\mu$ on the thermal fluctuations. The figures show the ``final" mean magnetization, obtained after 4000 time steps, as a function of the temperature $T$ (Figs.~\ref{effectmu1} and \ref{effectmu3}) and as a function of the effective or scaled temperature $\Theta$ (Figs.~\ref{effectmu2} and \ref{effectmu4}) for different values of $\mu$. All simulations are done on scale-free networks with 1000 nodes, $m=2$ and $\gamma = 3$. We used $a = 1.0$ and $b = 0.4$ in Figs.~\ref{effectmu1} and \ref{effectmu2}, while $a=0.85$ and $b = 0.4$ in Figs.~\ref{effectmu3} and \ref{effectmu4}. All data points are averages over 10 network realizations.
 \label{magnetvstmu}}
\end{figure}

Simulation results for different values of $\mu$ are shown in Fig.~\ref{effectmu1}, for $a=1$, and in Fig.~\ref{effectmu3} for $a\neq 1$. The magnetization curves tend to be stretched and shifted as $\mu$ decreases, reflecting the fact that networks remain stable up to a higher temperature for smaller values of $\mu$. Interestingly, for $a=1$, if we plot the magnetization as a function of the effective temperature $\Theta$, a good {\em data collapse} occurs, as is conspicuous in Fig.~\ref{effectmu2}. Not only do the different curves for $T > T_b$ in Figs.~\ref{effectmu1} fall onto a single curve in Fig.~\ref{effectmu2}, but also the different values of $T_b$ lead to practically one and the same value of $\Theta_b$. We argue that the high quality of the data collapse has two reasons. Firstly, for $a=1$ the scaling Ansatz \eqref{australie} is properly valid (at mean-field level) and secondly, the equilibrium critical temperatures for the different networks (in zero field) are not very different. (An exception could, in principle, have occurred in the borderline case $\gamma=3$ and $\mu=0$, for which $T_c$ diverges. But this divergence is very slow, in the manner $T_c \propto \log N$, for large $N$.) Notice how the quality of the data collapse degrades if $a\neq1 $, as can be seen by comparing Figs.~\ref{effectmu2} and~\ref{effectmu4}. In addition, there is now also a significant spread on the values of $\Theta_b$. These effects could have been  expected, since the scaling Ansatz \eqref{australie} is not well satisfied for $a \neq 1$, if $\mu$ is arbitrary.

We conclude that, at least in the range $0 < \mu < 1$, the effects of varying $\mu$ and varying $T$ are not independent. A change in $\mu$ can be compensated or ``absorbed" by a change in $T$. This is most clearly so for $a \approx 1$ and for $\gamma$ sufficiently large (i.e., in practice for $\gamma \geq 3$).

\subsubsection{Effect of the topological exponent $\gamma$}
In a second application of the use of the effective temperature of \eqref{theta}, we determine the effect of the network topology on the network activity or spin magnetization. We focus on the distribution model on scale-free networks with $m = 2$, 1000 nodes, constant $a$, $b$ and $\mu$ and we vary the topological exponent $\gamma$. Converting sums over the distribution function into integrals \cite{opm}, \eqref{theta} leads to
\begin{equation}
\Theta = \frac{\left(\gamma-2+\mu\right)^2}{(\gamma -1)(\gamma-2)}\frac{\left(K^{2-\gamma}-m^{2-\gamma}\right)\left(K^{1-\gamma}-m^{1-\gamma}\right)}{\left(K^{2-\gamma-\mu} - m^{2-\gamma-\mu}\right)^2}\:T .
\end{equation}
As in the previous subsection, we distinguish between effects at mean-field level and effects of the quenched random-bond fluctuations on the characteristic temperatures $T_b$ (breakdown) and $T_c$ (bulk criticality).

At mean-field level, the behavior of $\Theta$ as a function of $\gamma$ is complex and depends both on the parameter $\mu$ and on the size of the network through the maximal degree $K$. However, when $\mu = 0$ or $\mu > 1$, some simplifications apply. For $\mu = 0$, $\Theta \propto 1/\avq$ so that $\Theta$ increases with increasing $\gamma$, regardless of $m$ and $K$. For vanishing $\mu$, the network thus becomes more vulnerable to collapses if $\gamma$ increases, i.e., when there are less hubs in the network. In such a regime, hubs produce more goods than poorly connected nodes. The network thus appears more resilient against collapses when there are more hubs and when they are more productive.

The opposite behavior occurs for $\mu \geq 1$. For constant temperature $T$, $\Theta$ decreases with increasing $\gamma$ and thus the network becomes more vulnerable to collapses when $\gamma$ is decreased. The system is thus less prone to failure when there are less hubs and more nodes with a small degree. For large $\mu$, nodes with small degrees provide the larger quantities of goods. Between the two regimes, i.e., for $0<\mu<1$, there is a complex transition region in which the behavior of $\Theta$ as a function of $\gamma$ depends more subtly on the value of $\mu$ and the size of the network. Numerical simulations indicate that the effect of topology on the thermal fluctuations is rather small in this regime.

The effect of quenched random-bond disorder has already been discussed in the previous subsection. It suffices to recall that the value of the effective topological exponent $\gamma'$ determines whether we are dealing with a strongly expanded temperature scale ($T_c \rightarrow \infty$) or a normal one (with finite $T_c$).

\begin{figure}[t]
 \subfigure[$\mu$=0]{\includegraphics[angle=270,width = .5\textwidth]{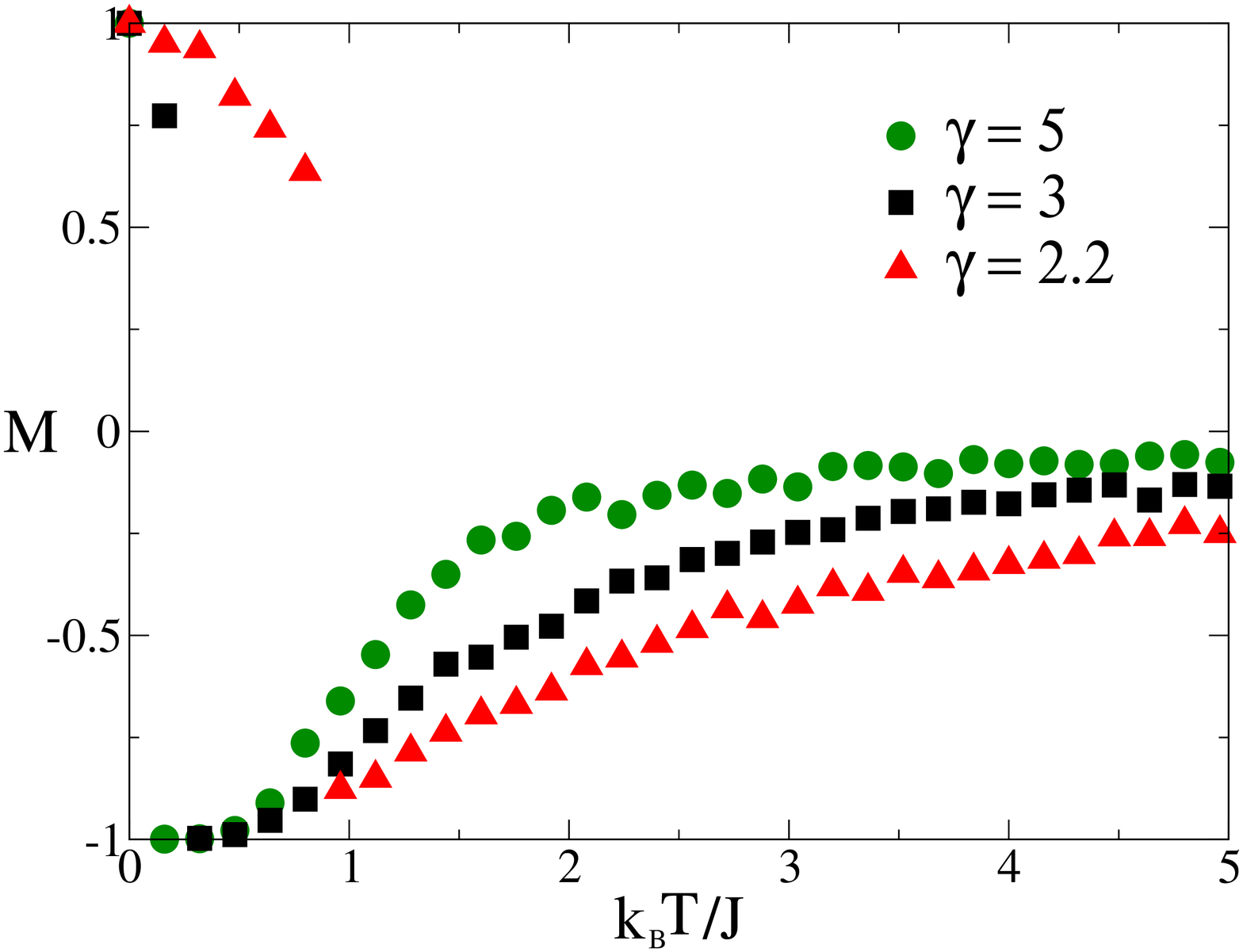}\label{5a}}
  \subfigure[$\mu$=0]{\includegraphics[angle=270,width = .5\textwidth]{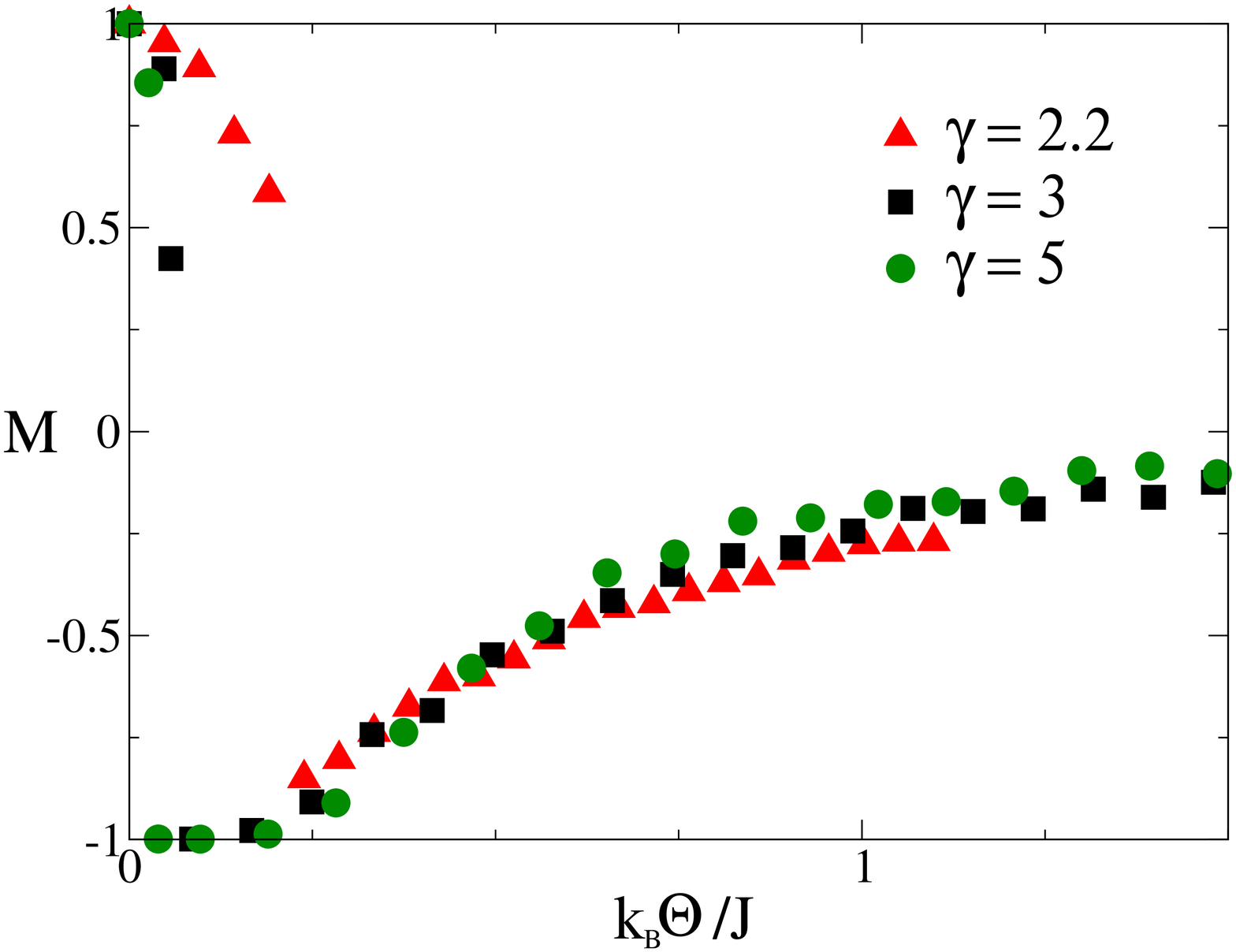}\label{5b}}\\
 \subfigure[$\mu$ = 1]{\includegraphics[angle=270,width = .5\textwidth]{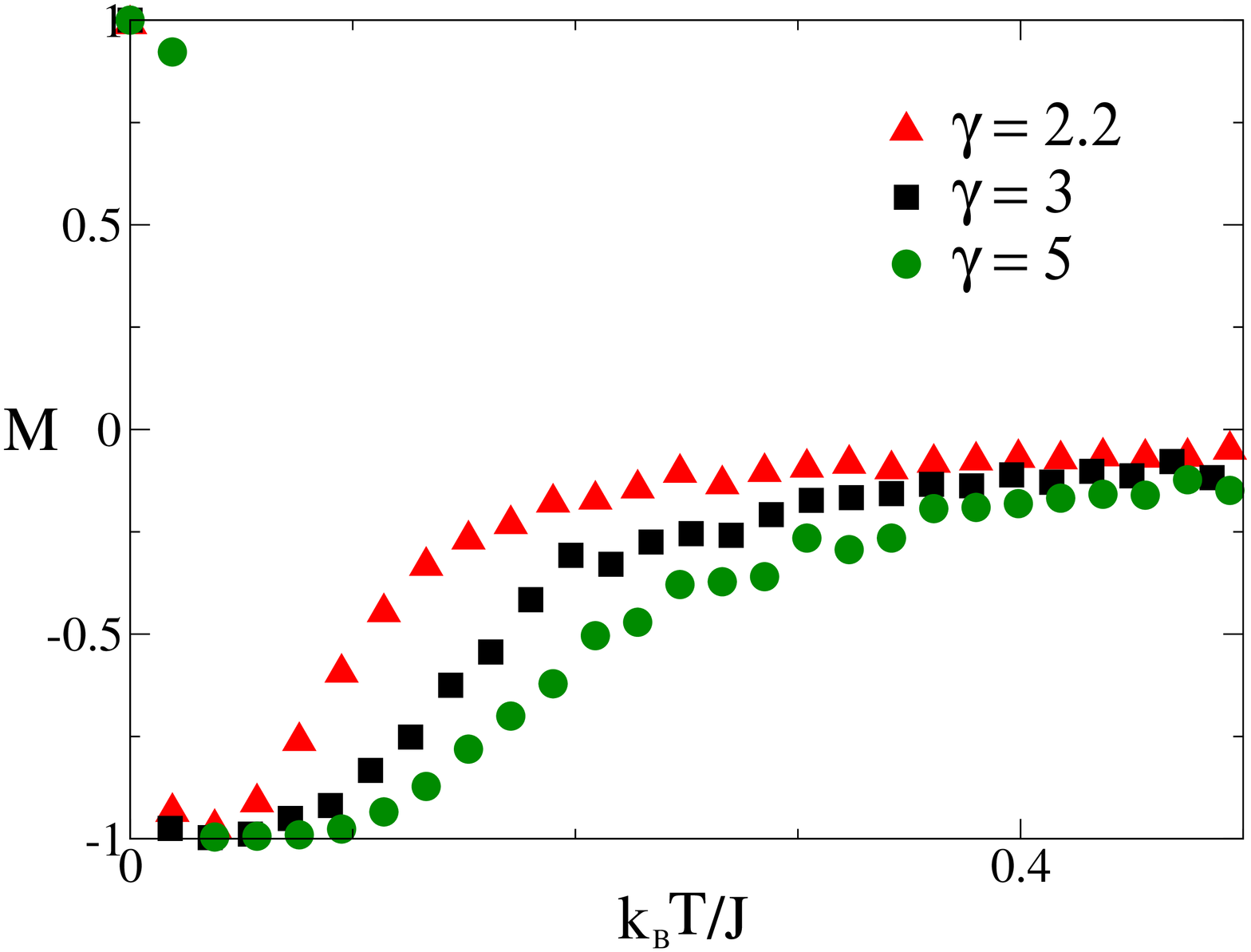}\label{5c}}
   \subfigure[$\mu$ =1]{\includegraphics[angle=270,width = .5\textwidth]{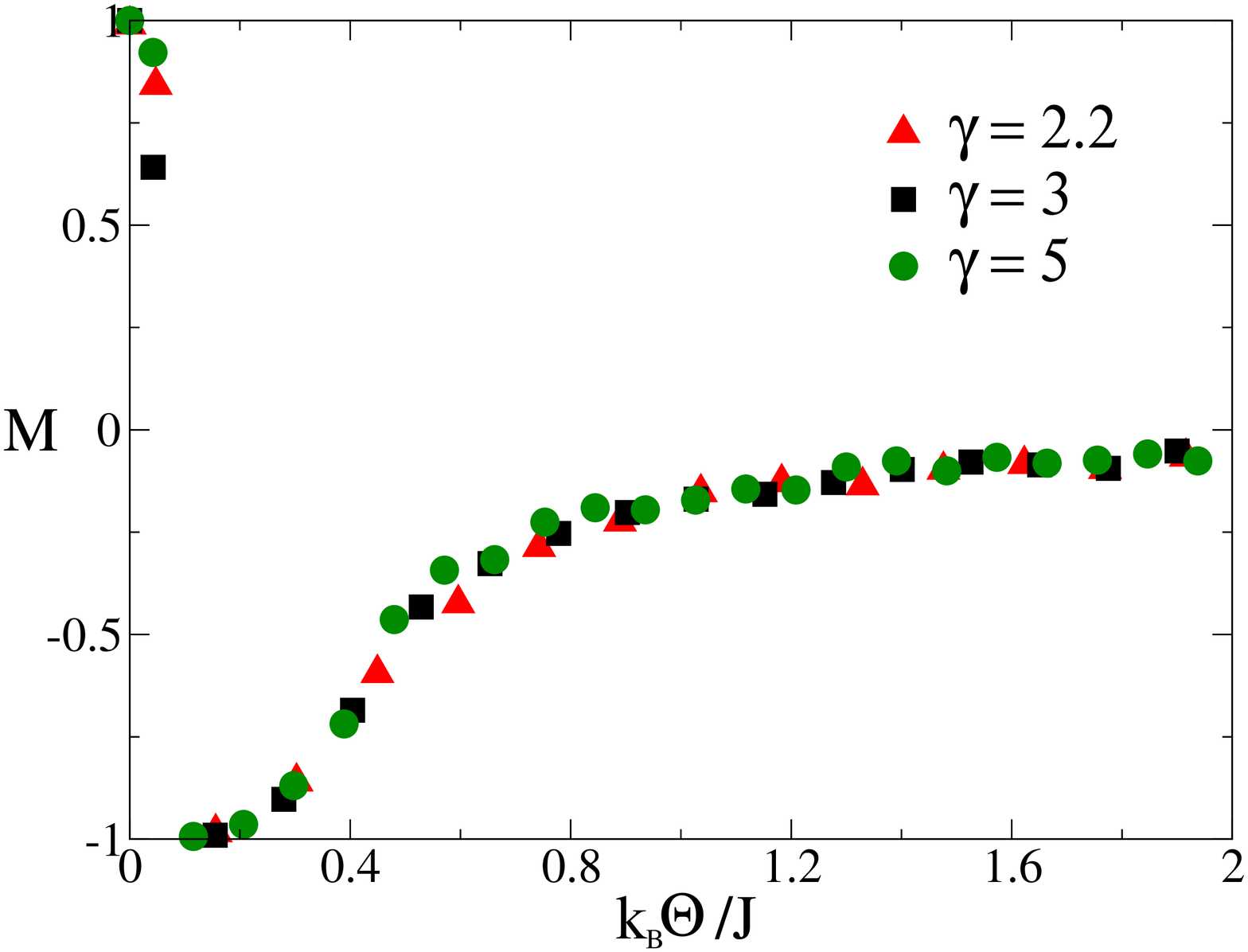}\label{5d}}
\caption{(Color online) Effect of the topology on the thermal fluctuations. The figures  show the magnetization as a function of the temperature $T$ (Figs.~(a) and (c)) and the parameter $\Theta$ (Figs.~(b) and (d)) for different values of $\gamma$. We study two different regimes depending on the parameter $\mu$. In Figs.~(a) and (b), $\mu = 0$, while $\mu = 1$ in Figs.~(c) and (d). All simulations were done on scale-free networks with 1000 nodes, with demand constants $a = 1.0$ and $b = 0.2$. All data points are averages over 10 network realizations.  \label{fig5}
}
\end{figure}

We illustrate the above qualitative features with simulation results in  Fig.~\ref{fig5}. In Fig.~\ref{5a}, the simulations use a distribution model with $\mu = 0$, while in Fig.~\ref{5c} $\mu = 1$ is taken. In both cases $a=1$ is assumed so that the mean-field Hamiltonian has the simple scaling form of \eqref{australie}. For $\mu = 0$ network collapses occur at lower temperatures for networks with larger $\gamma$, which implies that networks with more hubs and more highly connected hubs are more robust against thermal fluctuations. The opposite behavior is observed for $\mu = 1$ (see Fig.~\ref{5c}). The simulations thus confirm our expectations based on the dependence of $\Theta$ on $\mu$ and $\gamma$. In Figs.~\ref{5b} and \ref{5d}, the magnetization of the networks for the systems of Figs.~\ref{5a} and \ref{5c} is plotted as a function of the parameter $\Theta$.

In Fig.~\ref{5b} the data collapse is far from perfect. This is at first sight surprising because the conditions $a=1$ and $\mu = 0$ seem ideal prerequisites from the point of view of the validity of the Ansatz of \eqref{australie}. However, the networks examined are qualitatively different in the sense that for $\gamma \leq 3$ the degree fluctuations are important ($\averageq{2}$ diverges for $N\rightarrow\infty$), while for $\gamma > 3$ they are not. If we take into account the finite network size, we obtain the following estimates for the equilibrium critical temperatures in zero external field, using the analytic results of previous works \cite{giuraniuc1,dorogov}. For $\gamma = $ 5, 3, and 2.2 we find $k_BT_c/J \approx $ 0.60, 1.95 and 3.89, respectively, which spans a broad range. These values appear consistent with the behaviour of $M(T)$ (in non-zero external field) shown in Fig.~\ref{5a}. The simple ``mean-field" scaling underlying the definition of the effective temperature $\Theta$ is not quite sufficient to suppress the rather large effect of the degree fluctuations for low $\gamma$. This is why the magnetization curves and the breakdown temperatures show only a rather poor data collapse in
Fig.~\ref{5b}.

In contrast, a much better ``universality" is clearly emerging in Fig.~\ref{5d}, which is for systems with $\mu =1$. For these systems $\gamma' = \infty$ so that the networks all behave effectively as Poissonian networks, with a finite $T_c$ which scales in a simple manner with $\avq$. It is conspicuous in Fig.~\ref{5d} that both the $M(T)$ and the effective breakdown temperatures coincide well for all three cases.

\section{Conclusions}
In this Paper, we introduced a model to describe cooperative behavior in various real-life distribution systems. We specialized to networks in which a policy of competition (attempting to create new demand) or solidarity (reluctance to meet too high a demand) among suppliers takes effect so that there is a tendency towards maximizing the gap between delivery and demand. The resulting positive feedback mechanism can cause an initially active network to function during a certain time and then collapse to an inactive state. To implement this our  model utilizes an Ising-spin system with quenched random fields. The couplings are ferromagnetic in order to describe the positive feedback policies: spins align preferentially with their neighbours (up: active state; down: inactive state). Each node, of degree $q$, models a supplier which has a certain degree-dependent delivery controlled by an exponent $\mu$ and each link models (a number of) consumers with a certain demand, controlled by the amplitudes $a$ (supplier-adjusted demand) en $b$ (global demand). The degree distribution $P(q)$ is (mostly) of power-law type, pertaining to a scale-free network with topological exponent $\gamma$. The system is studied analytically, in part by using mean-field approximations, and numerically using Monte-Carlo simulations with the Metropolis updating rule.

Avalanches between an active and an inactive state are typically only observed if the inactive state has a lower total energy than the active state, and if a metastable active state is present initially. These conditions lead to restrictions on the demand parameters $a$ and $b$. If the demand is too large, the active state will immediately decay to the ground state and no metastable active state exists. If the demand is too small, the active state or some glassy state with lower activity is the ground state and no breakdowns can be observed. The region in which collapses can occur also depends on the topological exponent $\gamma$ of the network and on the delivery exponent $\mu$. If $|\mu|$ is increased, the region in which avalanches can occur shrinks gradually.

Random malfunction in the suppliers is modeled by a temperature parameter $T$. At finite temperatures, thermal fluctuations can cause network collapses, which are prevented at $T=0$ by the metastability of the active state. During a collapse the roles of the hubs and of the poorly connected nodes has been monitored separately. It has been possible to identify which type of nodes is responsible for initiating the breakdown as a function of the distribution and network parameters. The lowest temperature $T_b$ at which a network blackout takes place depends strongly on the different parameters in the model. Increasing this ``breakdown temperature" is of great interest in the protection of the distribution system. The most stable situation appears to be that in which there are many large suppliers in the network.  For large values of $\mu$, this requires many poorly connected nodes while for $\mu=0$ many hubs are needed. Rendering the amount of goods every consumer receives more homogeneous throughout the network also reduces the impact of thermal fluctuations in the system. Such a procedure can for instance be realized by decreasing $|\mu|$. Not all model parameters are independent. We have identified a scaled or effective temperature variable $\Theta$ which incorporates the $\gamma$ and $\mu$ dependencies in such a way that a good data collapse can occur when the network activity for various cases is measured as a function of $\Theta$ instead of $T$. While this scaling is very useful at the level of a mean-field approximation, we also observed that fluctuations in the node degree, which are large for networks with $\gamma \leq 3$, are responsible for deviations from simple scaling if also $\gamma' = (\gamma - \mu)/(1-\mu) \leq 3$.

The similarity between the model and real-life distribution systems can be improved in various ways. In the distribution model all the consumers depend on two suppliers. Using a bipartite network, i.e., a network with two kinds of nodes with links running only between unlike nodes, we can extend the model to incorporate an unrestricted number of suppliers for each consumer, reflecting the consumer's freedom of choice. Another extension is concerned with partly active suppliers. In the current model, a node is active or inactive. Using continuous spins, or discrete Potts spins, we could also model suppliers with tunable activity. Extensions of our model could also describe systems that evolve in time, for instance by implementing the model on growing networks or on networks in which rewiring is possible.  Time-dependent demand parameters $a$ and $b$ could be used to model the evolving economic characteristics of the consumers, etc.

We conclude that the distribution model introduced in this Paper offers a possible starting point for studying collapses in certain real-life distribution systems, from the viewpoint of the phenomenology of dynamical critical phenomena with a non-conserved order parameter. Note that in contrast with (most) other models of distribution or transportation networks \cite{Simonsen} there is no conservation law or continuity equation in our network. The amounts of goods flowing along the edges of the network are stochastic variables controlled by fluctuating spin states. In this sense our distribution network based on spin variables provides a complementary approach.
\vspace{1cm}

{\em Acknowledgements}

It is a pleasure and an honour to dedicate this paper to Professor David Sherrington on the occasion of his 70th birthday. We thank economist Marc Lambrecht of K.U.Leuven for his interest in and his comments on this model. H.H. and B.V.S. thank the FWO-Vlaanderen for support.

\newpage


\begin{thebibliography}{20}
\bibitem{barab1} A.L. Barab\'asi and R. Albert, Science {\bf 286}, 509 (1999).\\ A.L. Barab\'asi, H. Jeong and R. Albert, Nature {\bf
406}, 378 (2000).
\bibitem{strogatz} S.H. Strogatz, Nature {\bf 410}, 268 (2001).
\bibitem{newmanrev} M.E.J. Newman, SIAM Review  {\bf 45}, 167 (2003).
\bibitem{dorogov} S.N. Dorogovtsev, A.V. Goltsev and J.F.F. Mendes, Rev. Mod. Phys. {\bf 80}, 1275 (2008).
\bibitem{heiko} H. Bauke, C. Moore, J.-B. Rouquier and D. Sherrington, http://www.santafe.edu/media/workingpapers/11-05-017.pdf (2011).
\bibitem{indekeu1} J.O. Indekeu, Physica A {\bf 333}, 461 (2004).
\bibitem{thesisGiuraniuc} C.V. Giuraniuc, PhD-Thesis, K.U.Leuven (2006).
\bibitem{malarz} K. Malarz, J.Karpi\'nska, K.Kulakowski and
B.Tadi\'c, Physica A {\bf 373}, 785 (2007).
\bibitem{goh} K.-I. Goh, D.-S. Lee, B. Kahng and D. Kim, Phys. Rev. Lett. {\bf 91}, 148701 (2003).
\bibitem{sachtjen} M.L. Sachtjen, B.A. Carreras and V.E. Lynch, Phys. Rev. E {\bf 61}, 4877 (2000).
\bibitem{watts} D.J. Watts, PNAS {\bf 99}, 5766 (2002).
\bibitem{motter} A.E. Motter and Y.-C. Lai, Phys. Rev. E {\bf 66}, 065102(R) (2002).
\bibitem{crucitti} P. Crucitti, V. Latora, M. Marchiori, Phys. Rev. E {\bf 69} 045104 (2004).
\bibitem{buldyrev}S.V. Buldyrev, R. Parshani, G. Paul, H.E. Stanley and S. Havlin, Nature {\bf 464}, 1025 (2010).
\bibitem{giuraniuc1} C.V. Giuraniuc, J.P.L. Hatchett, J.O. Indekeu, M. Leone, I. Perez Castillo, B. Van Schaeybroeck and C. Vanderzande, Phys. Rev. Lett. {\bf 95}, 098701 (2005); ibid., Phys. Rev. E {\bf 74}, 036108 (2006).
\bibitem{gellings} C. W. Gellings and K. E. Yeager, Phys. Today {\bf
57}, 45 (2004).
\bibitem{cohen} R. Cohen, K.Erez, D. ben Avraham and S. Havlin, Phys. Rev. Lett. {\bf 85}, 4626 (2000).
\bibitem{catanzaro} M. Catanzaro, M. Boguna and R. Pastor-Satorras, Phys. Rev. E {\bf 71}, 027103 (2005).
\bibitem{thesisSVL} S. Van Lombeek, Master Thesis, K.U.Leuven (2010).
\bibitem{opm} A numerical calculation of $\Theta$ using the exact sums over the degree distribution yields similar results. There are only small numerical  differences.
\bibitem{Aleksiuk} A. Aleksiejuk, J.A. Holyst, D. Stauffer, Physica A {\bf 310}, 260 (2002).
\bibitem{Simonsen} I. Simonsen, L. Buzna, K. Peters, S. Bornholdt and D. Helbing, Phys. Rev. Lett. {\bf 100}, 218701 (2008).





\end{thebibliography}
\end{document}